\pdfoutput=1
\documentclass[superscriptaddress, 12pt, aps, prl,preprint,longbibliography]{revtex4-1}
\usepackage{graphicx}
\usepackage{epstopdf}
\usepackage{color}
\usepackage{amsmath, amssymb}
\usepackage{bm}
\usepackage{float}
\usepackage{placeins}
\usepackage{physics}
\usepackage{xr}

\begin{document}

\title{Dielectric constant of supercritical water in a large pressure-temperature range}
\author{Rui Hou}
\affiliation{Department of Physics, Hong Kong University of Science and Technology, Hong Kong, China}
\author{Yuhui Quan}
\affiliation{Department of Physics, Hong Kong University of Science and Technology, Hong Kong, China}
\author{Ding Pan}
\email{dingpan@ust.hk}
\affiliation{Department of Physics, Hong Kong University of Science and Technology, Hong Kong, China}
\affiliation{Department of Chemistry, Hong Kong University of Science and Technology, Hong Kong, China}
\affiliation{HKUST Fok Ying Tung Research Institute, Guangzhou, China}
\date{\today}

\begin{abstract}
 A huge amount of water at supercritical conditions exists in Earth's interior, where its dielectric properties play a critical role in determining how it stores and transports materials. 
 However, it is very challenging to obtain the static dielectric constant of water, $\epsilon_0$, in a wide pressure-temperature (P-T) range as found in deep Earth either experimentally or by first-principles simulations. Here, we introduce a neural network dipole model, which, combined with molecular dynamics, can be used to compute P-T dependent dielectric properties of water as accurately as first-principles methods but much more efficiently. 
 We found that $\epsilon_0$ may vary by one order of magnitude in Earth's upper mantle, suggesting that the solvation properties of water change dramatically at different depths. There is a subtle interplay between the molecular dipole moment and the dipolar angular correlation in governing the change of $\epsilon_0$. We also calculated the frequency-dependent dielectric constant of water in the microwave range, which, to the best of our knowledge, has not been calculated from first principles, and found that temperature affects the dielectric absorption more than pressure. 
 Our results are of great use in many areas, e.g., modelling water-rock interactions in geochemistry. 
 The computational approach introduced here can be readily applied to other molecular fluids.
\end{abstract}

\maketitle

\section{introduction}
Water is arguably the most important solvent on Earth, and also largely exists inside Earth's crust, mantle \cite{hirschmann2012water}, and even towards core \cite{mao2017water}. The supercritical water at pressure (P) above 22 MPa and temperature (T) higher than 647 K is a major component of Earth’s deep fluids \cite{liebscher2010aqueous}, storing and transporting many materials, and is also used as working fluids in various industrial areas \cite{weingartner2005supercritical}.
The dielectric constant of supercritical water, $\epsilon_0$, substantially affects its solvation properties and interactions with minerals, and is a key quantity in many science and industry applications \cite{fernandez1997formulation, weingartner2005supercritical}. 

For more than two decades, experimental data of the static dielectric constant of water has been limited to pressure lower than 0.5 GPa and temperature below 900 K \cite{fernandez1995database}, which can be extrapolated to $\sim$1 GPa and $\sim$1300 K using various models (e.g., \cite{fernandez1997formulation}). However, these P-T conditions can be only found in the very shallow mantle.
Experimentally, we do not know $\epsilon_0$ in most part of Earth's interior, where many important aqueous reactions are happening \cite{manning2018fluids}. First-principles molecular dynamics (FPMD) simulations have shown reliable predictions beyond the reach of current experiments, but the computational expense is so high that the previous study reported only 5 data points \cite{pan_dielectric_2013}. 
In molecular dynamics (MD) simulations, the calculation of $\epsilon_0$ requires the variance of the total dipole moment of the simulation box, $\vec{M}$, so a large number of uncorrelated configurations are needed \cite{neumann_computer_1984}. 
For example, at ambient conditions several nanoseconds MD simulations are required to get a converged value of $\epsilon_0$ \cite{gereben_accurate_2011}. 
In electronic structure calculations with periodic boundary conditions, 
$\vec{M}$ is often calculated by the sum of molecular dipole moments using maximally localized Wannier functions (MLWFs), obtained by minimizing the spread of molecular orbitals from  density functional theory (DFT) calculations \cite{marzari_maximally_2012}.

In recent years, machine learning techniques emerge as an appealing tool to combine the accuracy of first-principles simulations and the efficiency of empirical force fields in atomistic simulations.
In many studies, machine learning models were trained using data from first-principles calculations to learn potential energy surfaces, which can be further used in MD or Monte Carlo simulations (e.g., refs \cite{behler_generalized_2007, bartok_gaussian_2010,rupp_fast_2012,smith_ani-1_2017,zhang_deep_2018}).   
Some electronic properties may be also obtained using machine learning models, e.g.,
the dielectric constant of a variety of crystals \cite{umeda_prediction_2019}, but how the dielectric constant varies with environmental factors like P or T is not known.  

Here, we constructed a neural network dipole (NND) model using data from FPMD simulations to calculate the molecular dipole moment of supercritical water. In combination with MD trajectories obtained by the machine learning force field, we calculated the dielectric constant of water from 1 to 15 GPa and 800 to 1400 K, where $\epsilon_0$ changes by one order of magnitude. 
We also calculated the frequency-dependent dielectric constant of water, and found that temperature affects the dielectric absorption peak more than pressure.
The accurate and efficient method of calculating dielectric constant of water in a large P-T range makes it possible to model aqueous solutions and water-rock interactions in a large part of Earth's interior. Our results have great implications on the solvation properties of aqueous geofluids in deep Earth.

\section{Implementation details}
\begin{figure}
\centering
\vspace{5mm}
\includegraphics[width=0.6 \textwidth]{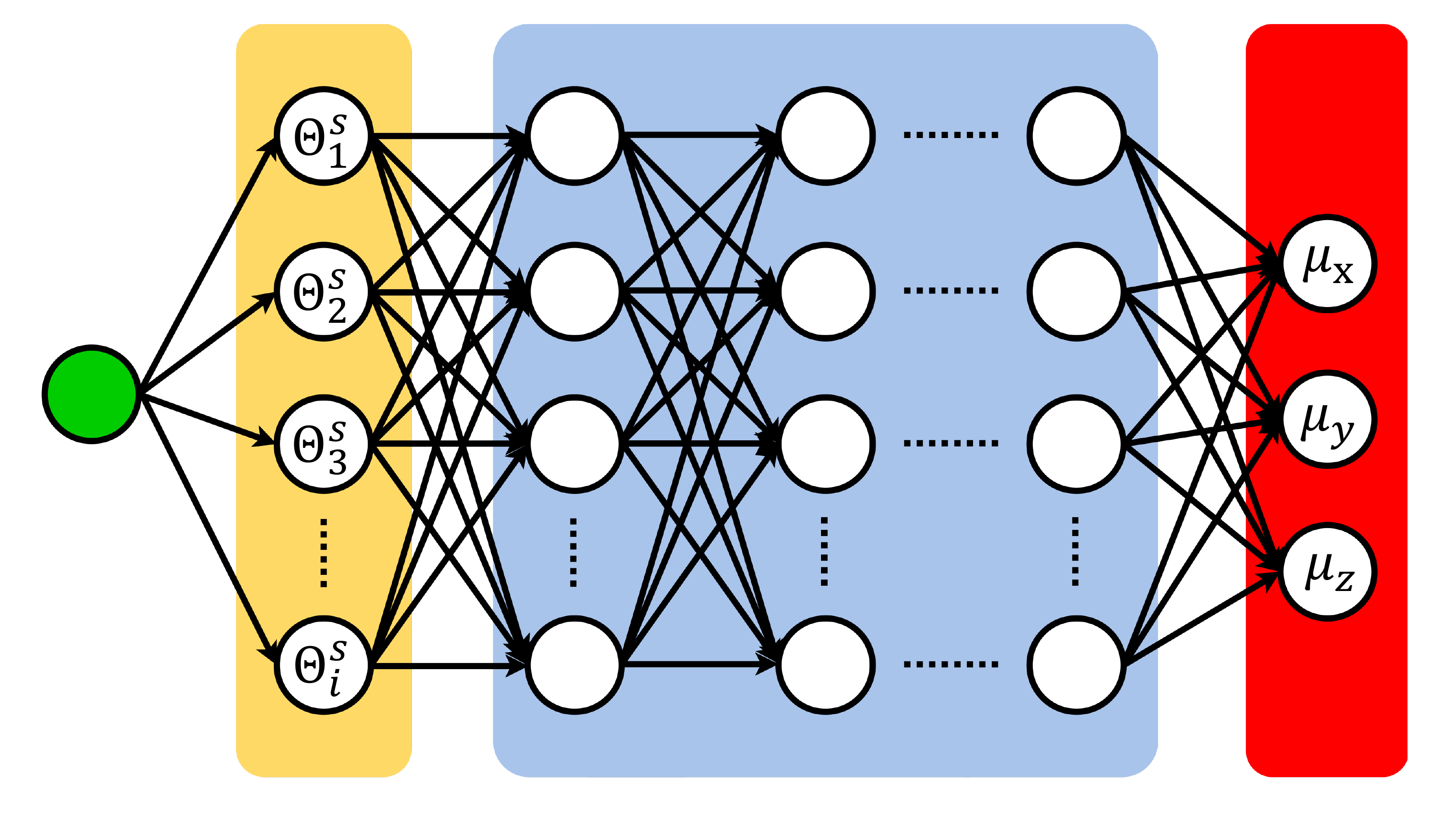}
\caption{Structure of the neural network used to output the molecular dipole moment. The green circle refers to the solvation shell structure as the initial data. The yellow, blue, and read blocks represent the input, hidden, and output layers, respectively. The molecular dipole moment, $\vec{\mu}$, is expressed using local coordinates, $(\mu_x, \mu_y, \mu_z)$.}
\label{fig:nn}
\end{figure}

In MD simulations with periodic boundary conditions, the dielectric constant of an isotropic and homogeneous fluid
can be calculated by \cite{neumann_computer_1984, sharma_dipolar_2007}:
\begin{equation} \label{epsilon_0}
  \begin{split}
    \epsilon_{0}=\epsilon_\infty+\frac{4 \pi}{3 k_{B} T V}\left(\left\langle\vec{M}^{2}\right\rangle-\langle\vec{M}\rangle^{2}\right),
  \end{split}
\end{equation}
where $\epsilon_\infty$ is the electronic dielectric constant, $k_B$ is the Boltzmann constant, $T$ is the temperature, $V$ is the volume of the simulation box. $\epsilon_\infty$ can be calculated by density functional perturbation theory \cite{baroni2001phonons}, whose fluctuation is much smaller than that of $\epsilon_{0}$, so tens of MD configurations are enough to get converged results \cite{pan2014refractive}.
The calculation of the variance of $\vec{M}$ requires a large number of uncorrelated MD configurations and the simulation box cannot be too small, so if we directly build a machine learning model to obtain $\vec{M}$ \cite{PhysRevLett.120.036002}, the training set has to sample the huge configuration space well, which can be very expensive. 
Instead, here we constructed the machine learning model
to get the molecular dipole moment of water in fluids. In DFT calculations, there are 64$\sim$128 water molecules in one simulation box, so we have 64$\sim$128 independent water dipoles from one MD configuration, which makes it easier to have a large training set. The dipole moment of water molecules 
strongly depends on their local chemical environment \cite{pan_dielectric_2013}, so 
we chose the local solvation structure as input to an end-to-end neural network model
as shown in Fig. \ref{fig:nn}. We built a local Cartesian coordinate for each water molecule based on its molecular geometry. 
The origin is located at the O atom of the center water molecule. The $x$ axis is along the bisector of $\angle$HOH, and the $z$ axis is perpendicular to the plane formed by one O and two H atoms.
In this way, the rotational and translational symmetries are preserved. For the center water molecule, the input is the coordinates of its neighboring atoms including its two H atoms $\vec{\Theta}=\{\Theta^O_1, \Theta^O_2,...,\Theta^O_i,...; \Theta^H_1, \Theta^H_2,...,\Theta^H_i,... \}$  within a cutoff radius R$_c$:
\begin{equation}
    \Theta^s_{i} = \{\frac{x_i}{r_{i}^3}, \frac{y_i}{r_{i}^3}, \frac{z_i}{r_{i}^3}\}, r_{i} \leq R_c,
\end{equation}
where $(x_i,y_i,z_i)$ is the local Cartesian coordinate of the $i$th neighboring atom, $s$ labels the species as either O or H, and $r_i = \sqrt{x_i^2+y_i^2+z_i^2}$.
We sorted the atoms by distance from near to far in the input to preserve the permutational symmetry.
We fixed the cutoff radius at 0.6 nm, and $\vec{\Theta}$ has the coordinates of 32 water molecules. In our study, the water density varies considerably, so if the number of water molecules within this radius was smaller than 32, we appended (0,0,0) to the tail of $\vec{\Theta}$ to keep the size of $\vec{\Theta}$ constant.

In the neural network model, we used four hidden layers, sequentially consisting of 300, 200, 100, and 30 nodes, to connect the input and output layers. We chose the hyperbolic tangent function \cite{tensorflow2015-whitepaper} 
as the activation function for all hidden layers. The loss function is defined as,
\begin{equation} \label{eq1}
    \begin{split}
        L=\frac{1}{n}\sum_{j=1}^n [(\vec{\mu}^p_j-\vec{\mu}^t_j)^2 + (\norm{\vec{\mu}^p_j}-\norm{\vec{\mu}^t_j})^2]
    \end{split}
\end{equation}
where $\vec{\mu}^p_j$ and $\vec{\mu}^t_j$ are
the predicted and inputted dipole moments of the $i$th water molecule, respectively, and $n$ is the batch size.
We used the Adam optimization algorithm \cite{kingma_adam_2017} to minimize the loss function and the learning rate is 1.0 $\times 10^{-4}$. To avoid overfitting,  we added a L2 regularization \cite{10.1145/1015330.1015435} to each hidden layer. 

Our training set consists of 2138880 water dipoles obtained from over 100 ps FPMD simulations at ambient and supercritical conditions up to 11 GPa and 2000 K. Although our current study is focused on supercritical water, we found that the data at ambient conditions could improve the accuracy of the trained NND model. 

To calculate the dipole moment of water molecules, we used the PBE xc functional \cite{Perdew1996}, which may not be sufficient enough to describe hydrogen bonds in water at ambient conditions \cite{zhang2011first}, but our previous studies indicated that it performs better at high P and high T than at ambient conditions \cite{Pan2016}, particularly for dielectric properties \cite{pan_dielectric_2013, pan2014refractive}. 

\section{results and discussion}

 We compared the distributions of dipole magnitude of water molecules obtained by the NND model and DFT in the test set (see Fig. S1), where the mean absolute error (MAE) is 0.14 D. Interesting, we also tested our NND model at 30 GPa and 2000 K, which was not included in the training set, and the MAE is 0.038 D. 
 The water ionization happens very frequently at this condition, so we use one O atom and two nearest H atoms to form a water `molecule' to calculate the molecular dipole. The excellent performance at 30 GPa and 2000 K suggests that our NND model can not only cover the P-T range of the training set, but may also be extrapolated outside this range.

In Fig. S2 we plotted the distributions of the angle between the dipole moment vectors calculated by the NND model and DFT, and the MAE is 3.86$^{\circ}$ for individual dipole moments. It is interesting to see that the error of $\vec{M}$ is slightly smaller than that of individual dipoles, which may be due to the error cancellation in the vector summation.

\begin{figure}
\centering
\vspace{5mm}
\includegraphics[width=0.6\textwidth]{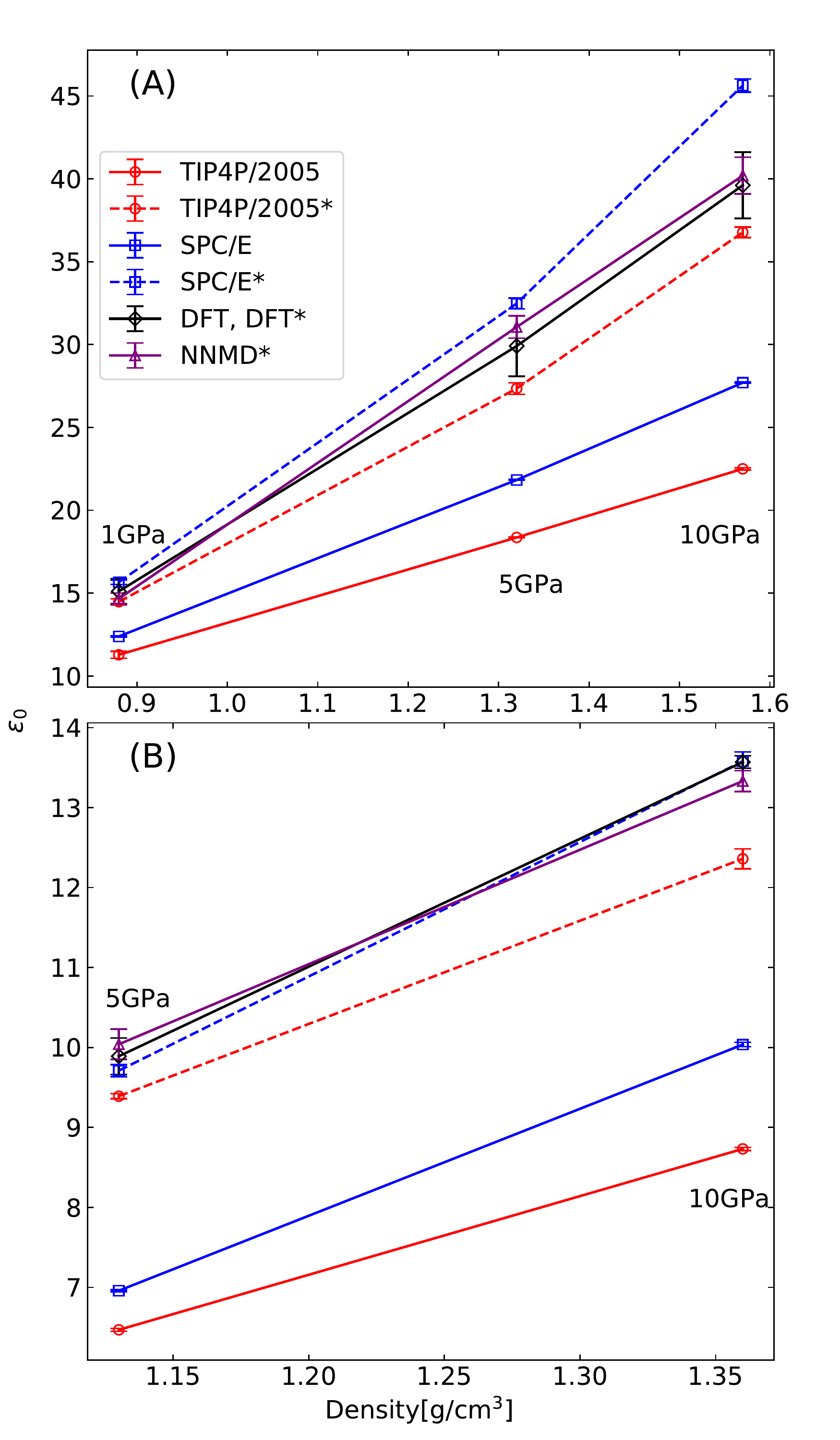}
\caption{Static dielectric constant of water as a function of density. Two temperatures are compared: (A) 1000 K and (B) 2000 K. Results are obtained from the MD trajectories using the DFT, neural network (NNMD), and empirical force fields (SPC/E, TIP4P/2005). The asterisk (*) denotes that the trajectory is from the corresponding force field, while molecular dipoles are calculated using the neural network dipole model. The DFT and DFT* results are indistinguishable. Error bars are obtained by the blocking method \cite{flyvbjerg1989error}.}
\label{fig_validate}
\end{figure}

After validating water dipole moments obtained by our NND model, we applied them to calculate  $\epsilon_0$ at various P-T conditions as shown in Fig. \ref{fig_validate}. 
At each P-T condition, we used half of the MD trajectory to train the NND model and the remaining half to assess results.
Overall, the relative error is smaller than 0.6\% in test sets, which is much smaller than the standard deviations obtained from our MD simulations, indicating that our NND model is nearly as accurate as first principles calculations (see Table SI).

After building an excellent model to calculate the molecular dipole moment of water, we may conduct first-principles MD simulations to generate trajectories at various P-T conditions and then use the NND model to calculate the fluctuation of $\vec{M}$ to get the dielectric constant of water.
Using this method, we avoid the step of calculating MLWFs.
However, the FPMD simulation is also very expensive, which limits its application to many P-T conditions. 
In many previous studies on high P-T water, the force fields SPC/E \cite{berendsen_missing_1987} and TIP4P/2005 \cite{doi:10.1063/1.2121687} were widely used. Although they were mainly designed to simulate ambient water, they seem to work well for some properties at high P and T, e.g., the equation of state \cite{zhang_prediction_2005}. As for dielectric properties, the main limitation of these two models is their rigidity. The molecular dipole moment of SPC/E is fixed at 2.34 D, while that of TIP4P/2005 is 2.31 D. In fact, both P and T may affect the molecular dipole moment of water considerably.
Our previous study shows that at $\sim$1 GPa and 1000 K, the dielectric constant of water calculated by the SPC/E model is close to the result obtained from FPMD simulations and also the extrapolated experimental value \cite{pan_dielectric_2013}; however, the difference between the SPC/E result and that from FPMD simulations becomes larger with increasing pressure. We found that at $\sim$1 GPa and 1000 K, the average dipole moment of water molecules,  $\mu$, calculated by DFT is close to 2.34 D, but $\mu$ increases with increasing pressure and decreases with increasing temperature, which cannot be captured by rigid water models. 

Since our NND model gives excellent dipole distributions under various P-T conditions, we applied this model to calculate molecular dipole moments for the trajectories obtained from the SPC/E and TIP4P/2005 simulations, as shown in Fig. \ref{fig_validate}.
The dielectric constants calculated by our NND model are generally larger than those directly from the two water models, and become closer to the DFT values, indicating that the polarization of water molecules with pressure and temperature affects the dielectric constant of water.

It is interesting to see that the dielectric constants calculated by the SPC/E model are closer to the DFT values than those by the TIP4P/2005 model, whereas at 1000 K using the NND model
the TIP4P/2005 trajectories give better results than the SPC/E ones. 
The dielectric constant of water depends on two factors: (1) molecular dipole moment and (2) hydrogen bond network characterized by the Kirkwood g-factor \cite{kirkwood1939dielectric}:
\begin{equation}
    G_k=\frac{\langle \vec{M}^2\rangle}{N\mu^2},
\end{equation}
where $N$ is the number of molecules in the simulation box.
Vega et al. argued that for high pressure ice phases at 243 K, TIP4P/2005 may provide decent $G_k$, so after scaling the molecular dipole moment, it gives correct dielectric constants \cite{aragones2011dielectric}, which is qualitatively consistent with our finding at 1000 K. 
However, when increasing T to 2000 K, 
the SPC/E trajectories perform better again
than TIP4P/2005, 
indicating that the SPC/E model may give more 
accurate $G_k$ than TIP4P/2005 at very high temperature.

The variation of molecular geometry of water can not be well reproduced by rigid force fields. Here, we also generated MD trajectories using the neural network force field, which was recently developed for high P-T water and ice in the molecular, ionic, and superionic phases up to 70 GPa and 3000 K using the DFT energy potential \cite{LinZhuang-43101}. We
calculated $\epsilon_0$ using the obtained water trajectories, as shown in Fig. \ref{fig_validate}. 
The difference between the NND and DFT values is smaller than 1.7\% , which is within error bars in FPMD simulations and overall better than the results obtained from the SPC/E and TIP4P/2005 trajectories.

\begin{figure}
\centering
\vspace{5mm}
\includegraphics[width=0.6 \textwidth]{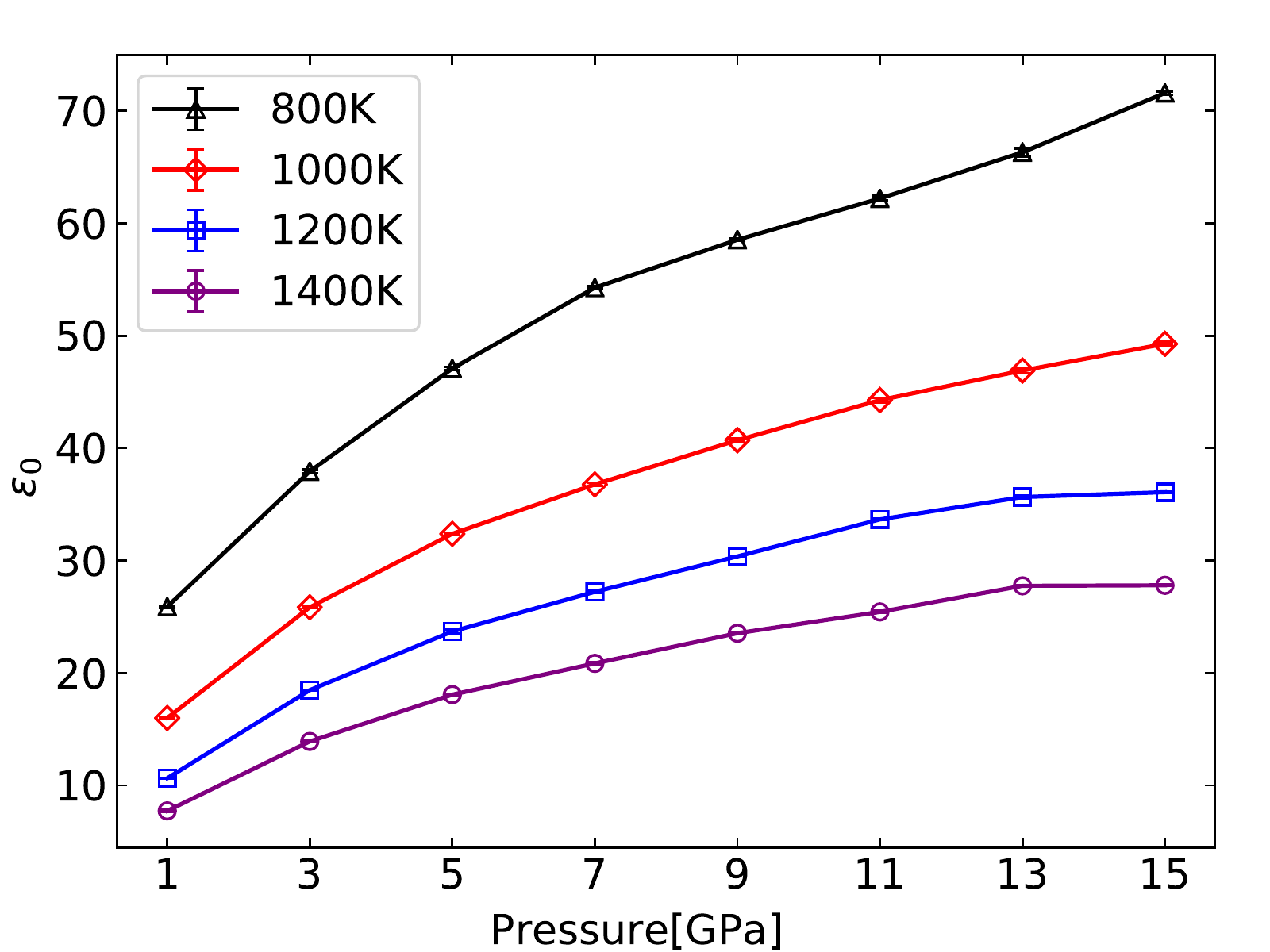}
\caption{Static dielectric constant of supercritical water at high P-T conditions.
Molecular dipoles are calculated using the neural network dipole model and trajectories are from simulations with the neural network force field. Error bars are within symbols. The equation of state of water is from Ref. \cite{zhang_prediction_2005}.}
\label{fig:eps0}
\end{figure}

Using the NND model and the neural network force field, we calculated the dielectric constant of supercritical water from 1 to 15 GPa and 800 to 1400 K, corresponding to the P-T conditions found in Earth's upper mantle.
$\epsilon_0$ increases with increasing P and decreases with increasing T, and varies from $\sim$7 up to 72 in Fig. \ref{fig:eps0}. 
Generally, the increase of $\epsilon_0$ becomes smaller with increasing P along an isotherm. 
Since $\epsilon_0$ determines the solvation properties of water, 
the large variation of $\epsilon_0$ substantially 
influences how water stores and transports materials with great implications on water-rock interactions in Earth's interior. 

Fig. 4 shows $\mu$ and $G_k$ as a function of pressure.
With increasing P along an isotherm, $\mu$ increases monotonically, whereas $G_k$ increases and then decreases; its peak appears at 5$\sim$7 GPa. 
$G_k$ accounts for the dipolar angular correlation.
With increasing P from 1 to 5 GPa, the angular correlation among dipoles is enhanced because the molecular interaction becomes stronger. With increasing P further, however, the hydrogen bonding becomes weaker due to the increased coordination number \cite{schwegler2000water}, so the angular correlation is not as strong as at low P.
There is a subtle interplay between $\mu$ and $G_k$ in governing the change of $\epsilon_0$.
The increase of $\epsilon_0$ at low P is driven by the increase of $\mu$, $G_k$, and water density, while at high P it is because the increase of $\mu$ and density exceeds the decrease of $G_k$; as a result, the increase of $\epsilon_0$ at high P is generally smaller than that at low P.

The dielectric constant is a function of the electric field frequency,  $\epsilon(\omega)$, which can be calculated  by the Fourier-Laplace transform of the derivative of the normalized autocorrelation function of $\vec{M}$, $ \Phi(t) = \frac{<\vec{M}(0)\cdot\vec{M}(t)>}{<\vec{M}^2>}$ \cite{neumann_computer_1984}:
\begin{align}
    \frac{\epsilon(\omega)-\epsilon_\infty}{\epsilon_0-\epsilon_\infty} &= \int_0^{\infty}(-\frac{d\Phi(t)}{dt})e^{-i\omega t}dt. \label{FLT}
\end{align}
It is difficult to get converged $\epsilon(\omega)$ in the microwave range from FPMD simulations.
Using the NND model and the neural network force field, we found that the rapid decay part of $\Phi(t)$ can be well fitted to an exponential function $A e^{-\frac{t}{\tau_D}}$, where $\tau_D$ approximates the Debye relaxation time \cite{van_der_spoel_systematic_1998}, with fitting errors smaller than 2.3\%. 
In our calculations, $\Phi(t)$ contains the raw data until it decreases to 0.2, the tail from the exponential function after $\Phi(t)$ is smaller than 0.1, and the linear interpolation between the raw and fitted data when $\Phi(t)$ is between 0.2 and 0.1 to make a smooth connection.
Fig. \ref{fig:epsw} shows the real and imaginary parts of the frequency-dependent dielectric constant of supercritical water in the microwave range.
The large peaks in Fig. \ref{fig:epsw}(B) correspond to the main dielectric absorption peaks, which are centered between about 600 GHz and 10 THz, much larger than that at ambient conditions, $\sim$ 20 GHz. 
The main absorption peak is upshifted with increasing temperature, but downshifted with increasing pressure.
The rise of temperature from 1000 to 2000 K affects the peak position more than the increase of pressure from 1 to 10 GPa.
The dielectric absorption of water may be attributed to single molecule or large collection motions, which is still under debate \cite{elton_origin_2017}.
For supercritical water studied here, it seems temperature has more obvious effects on the Debye relaxation time than pressure.

Note that very recently Zhang et al. introduced a neural network model to obtain the position of MLWF centers \cite{zhang2019weinan}, which can be also used to calculate molecular dipole moments, but this method may be not as efficient as the NND model introduced here, considering that we need to calculate four MLWF center positions in one water molecule.

\section{Conclusion}
In summary, we built a neural network dipole model, which combines the accuracy of first principles calculations and the efficiency of empirical models, to compute the dielectric constant of supercritical water,  $\epsilon_0$, from 1 to 15 GPa and 800 to 1400 K. We found that  $\epsilon_0$ can vary by one order of magnitude in Earth's upper mantle, suggesting that solvation properties of water may change dramatically at different depths, so water, as an important mass transfer medium, may dissolve some materials, e.g., carbon \cite{hazen2013deep}, and transport and release them in some shallow areas, which connects material reservoirs in Earth's surface and interior. A subtle interplay between the molecular dipole moment and the dipolar angular correlation governs the increase of $\epsilon_0$ with pressure.
We also calculated the dielectric constant as a function of the electric field frequency, and found that temperature affects the dielectric absorption peak more than pressure in the P-T range studied here. The accuracy of our method solely depends on the quality of the training data, which can be further improved by using high level theories, e.g., hybrid exchange-correlation functionals in DFT.  
Although we studied only water here, our method can be readily applied to other molecular fluids. 

The dielectric constant of water as a function of P and T plays a key role in the Deep Earth Water (DEW) model developed recently, which can calculate thermodynamic properties of many aqueous species and study water-rock interactions at elevated P-T conditions \cite{huang2019extended}.  
Using the method introduced here, we are able to build a high quality database for $\epsilon_0$ in a large P-T range as found in deep Earth.
The obtained data has great implications on how water stores and transports materials in Earth's interior.

\section{Acknowledgments}
We thank Giulia Galli and Viktor Rozsa for their helpful discussions, and Lin Zhuang and Xin-Zheng Li for helping the simulations with the neural network force field.
D.P. acknowledges support from the Croucher Foundation through the Croucher Innovation Award, Hong Kong Research Grants Council (Projects ECS-26305017 and GRF-16307618), National Natural Science Foundation of China (Project 11774072), and the Alfred P. Sloan Foundation through the Deep Carbon Observatory.
Part of this work was carried out using computational resources from the National Supercomputer Center in Guangzhou, China.

\section{Author contributions}
D.P. designed research; R.H., Y.Q, and D.P. performed research;
R.H., Y.Q, and D.P. analyzed data;
and R.H. and D.P. wrote the paper.

\bibliography{ref}

\begin{thebibliography}{40}%
\makeatletter
\providecommand \@ifxundefined [1]{%
 \@ifx{#1\undefined}
}%
\providecommand \@ifnum [1]{%
 \ifnum #1\expandafter \@firstoftwo
 \else \expandafter \@secondoftwo
 \fi
}%
\providecommand \@ifx [1]{%
 \ifx #1\expandafter \@firstoftwo
 \else \expandafter \@secondoftwo
 \fi
}%
\providecommand \natexlab [1]{#1}%
\providecommand \enquote  [1]{``#1''}%
\providecommand \bibnamefont  [1]{#1}%
\providecommand \bibfnamefont [1]{#1}%
\providecommand \citenamefont [1]{#1}%
\providecommand \href@noop [0]{\@secondoftwo}%
\providecommand \href [0]{\begingroup \@sanitize@url \@href}%
\providecommand \@href[1]{\@@startlink{#1}\@@href}%
\providecommand \@@href[1]{\endgroup#1\@@endlink}%
\providecommand \@sanitize@url [0]{\catcode `\\12\catcode `\$12\catcode
  `\&12\catcode `\#12\catcode `\^12\catcode `\_12\catcode `\%12\relax}%
\providecommand \@@startlink[1]{}%
\providecommand \@@endlink[0]{}%
\providecommand \url  [0]{\begingroup\@sanitize@url \@url }%
\providecommand \@url [1]{\endgroup\@href {#1}{\urlprefix }}%
\providecommand \urlprefix  [0]{URL }%
\providecommand \Eprint [0]{\href }%
\providecommand \doibase [0]{http://dx.doi.org/}%
\providecommand \selectlanguage [0]{\@gobble}%
\providecommand \bibinfo  [0]{\@secondoftwo}%
\providecommand \bibfield  [0]{\@secondoftwo}%
\providecommand \translation [1]{[#1]}%
\providecommand \BibitemOpen [0]{}%
\providecommand \bibitemStop [0]{}%
\providecommand \bibitemNoStop [0]{.\EOS\space}%
\providecommand \EOS [0]{\spacefactor3000\relax}%
\providecommand \BibitemShut  [1]{\csname bibitem#1\endcsname}%
\let\auto@bib@innerbib\@empty
%</preamble>
\bibitem [{\citenamefont {Hirschmann}\ and\ \citenamefont
  {Kohlstedt}(2012)}]{hirschmann2012water}%
  \BibitemOpen
  \bibfield  {author} {\bibinfo {author} {\bibfnamefont {Marc}\ \bibnamefont
  {Hirschmann}}\ and\ \bibinfo {author} {\bibfnamefont {David}\ \bibnamefont
  {Kohlstedt}},\ }\bibfield  {title} {\enquote {\bibinfo {title} {Water in
  {Earth’s} mantle},}\ }\href@noop {} {\bibfield  {journal} {\bibinfo
  {journal} {Phys Today}\ }\textbf {\bibinfo {volume} {65}},\ \bibinfo {pages}
  {40} (\bibinfo {year} {2012})}\BibitemShut {NoStop}%
\bibitem [{\citenamefont {Mao}\ \emph {et~al.}(2017)\citenamefont {Mao},
  \citenamefont {Hu}, \citenamefont {Yang}, \citenamefont {Liu}, \citenamefont
  {Kim}, \citenamefont {Meng}, \citenamefont {Zhang}, \citenamefont
  {Prakapenka}, \citenamefont {Yang},\ and\ \citenamefont
  {Mao}}]{mao2017water}%
  \BibitemOpen
  \bibfield  {author} {\bibinfo {author} {\bibfnamefont {Ho-Kwang}\
  \bibnamefont {Mao}}, \bibinfo {author} {\bibfnamefont {Qingyang}\
  \bibnamefont {Hu}}, \bibinfo {author} {\bibfnamefont {Liuxiang}\ \bibnamefont
  {Yang}}, \bibinfo {author} {\bibfnamefont {Jin}\ \bibnamefont {Liu}},
  \bibinfo {author} {\bibfnamefont {Duck~Young}\ \bibnamefont {Kim}}, \bibinfo
  {author} {\bibfnamefont {Yue}\ \bibnamefont {Meng}}, \bibinfo {author}
  {\bibfnamefont {Li}~\bibnamefont {Zhang}}, \bibinfo {author} {\bibfnamefont
  {Vitali~B}\ \bibnamefont {Prakapenka}}, \bibinfo {author} {\bibfnamefont
  {Wenge}\ \bibnamefont {Yang}}, \ and\ \bibinfo {author} {\bibfnamefont
  {Wendy~L}\ \bibnamefont {Mao}},\ }\bibfield  {title} {\enquote {\bibinfo
  {title} {When water meets iron at {E}arth's core--mantle boundary},}\
  }\href@noop {} {\bibfield  {journal} {\bibinfo  {journal} {Natl Sci Rev}\
  }\textbf {\bibinfo {volume} {4}},\ \bibinfo {pages} {870--878} (\bibinfo
  {year} {2017})}\BibitemShut {NoStop}%
\bibitem [{\citenamefont {Liebscher}(2010)}]{liebscher2010aqueous}%
  \BibitemOpen
  \bibfield  {author} {\bibinfo {author} {\bibfnamefont {Axel}\ \bibnamefont
  {Liebscher}},\ }\bibfield  {title} {\enquote {\bibinfo {title} {Aqueous
  fluids at elevated pressure and temperature},}\ }\href@noop {} {\bibfield
  {journal} {\bibinfo  {journal} {Geofluids}\ }\textbf {\bibinfo {volume}
  {10}},\ \bibinfo {pages} {3--19} (\bibinfo {year} {2010})}\BibitemShut
  {NoStop}%
\bibitem [{\citenamefont {Weing{\"a}rtner}\ and\ \citenamefont
  {Franck}(2005)}]{weingartner2005supercritical}%
  \BibitemOpen
  \bibfield  {author} {\bibinfo {author} {\bibfnamefont {Hermann}\ \bibnamefont
  {Weing{\"a}rtner}}\ and\ \bibinfo {author} {\bibfnamefont {Ernst~Ulrich}\
  \bibnamefont {Franck}},\ }\bibfield  {title} {\enquote {\bibinfo {title}
  {Supercritical water as a solvent},}\ }\href@noop {} {\bibfield  {journal}
  {\bibinfo  {journal} {Angew Chem Int Ed Engl}\ }\textbf {\bibinfo {volume}
  {44}},\ \bibinfo {pages} {2672--2692} (\bibinfo {year} {2005})}\BibitemShut
  {NoStop}%
\bibitem [{\citenamefont {Fernandez}\ \emph {et~al.}(1997)\citenamefont
  {Fernandez}, \citenamefont {Goodwin}, \citenamefont {Lemmon}, \citenamefont
  {Levelt~Sengers},\ and\ \citenamefont {Williams}}]{fernandez1997formulation}%
  \BibitemOpen
  \bibfield  {author} {\bibinfo {author} {\bibfnamefont {DP}~\bibnamefont
  {Fernandez}}, \bibinfo {author} {\bibfnamefont {ARH}\ \bibnamefont
  {Goodwin}}, \bibinfo {author} {\bibfnamefont {Eric~W}\ \bibnamefont
  {Lemmon}}, \bibinfo {author} {\bibfnamefont {JMH}\ \bibnamefont
  {Levelt~Sengers}}, \ and\ \bibinfo {author} {\bibfnamefont {RC}~\bibnamefont
  {Williams}},\ }\bibfield  {title} {\enquote {\bibinfo {title} {A formulation
  for the static permittivity of water and steam at temperatures from 238 k to
  873 k at pressures up to 1200 mpa, including derivatives and
  {D}ebye--{H}{\"u}ckel coefficients},}\ }\href@noop {} {\bibfield  {journal}
  {\bibinfo  {journal} {J Phys Chem Ref Data}\ }\textbf {\bibinfo {volume}
  {26}},\ \bibinfo {pages} {1125--1166} (\bibinfo {year} {1997})}\BibitemShut
  {NoStop}%
\bibitem [{\citenamefont {Fernandez}\ \emph {et~al.}(1995)\citenamefont
  {Fernandez}, \citenamefont {Mulev}, \citenamefont {Goodwin},\ and\
  \citenamefont {Sengers}}]{fernandez1995database}%
  \BibitemOpen
  \bibfield  {author} {\bibinfo {author} {\bibfnamefont {Diego~P}\ \bibnamefont
  {Fernandez}}, \bibinfo {author} {\bibfnamefont {Y}~\bibnamefont {Mulev}},
  \bibinfo {author} {\bibfnamefont {ARH}\ \bibnamefont {Goodwin}}, \ and\
  \bibinfo {author} {\bibfnamefont {JMH~Levelt}\ \bibnamefont {Sengers}},\
  }\bibfield  {title} {\enquote {\bibinfo {title} {A database for the static
  dielectric constant of water and steam},}\ }\href@noop {} {\bibfield
  {journal} {\bibinfo  {journal} {J Phys Chem Ref Data}\ }\textbf {\bibinfo
  {volume} {24}},\ \bibinfo {pages} {33--70} (\bibinfo {year}
  {1995})}\BibitemShut {NoStop}%
\bibitem [{\citenamefont {Manning}(2018)}]{manning2018fluids}%
  \BibitemOpen
  \bibfield  {author} {\bibinfo {author} {\bibfnamefont {Craig~E.}\
  \bibnamefont {Manning}},\ }\bibfield  {title} {\enquote {\bibinfo {title}
  {Fluids of the lower crust: Deep is different},}\ }\href {\doibase
  10.1146/annurev-earth-060614-105224} {\bibfield  {journal} {\bibinfo
  {journal} {Annu Rev Earth Planet Sci}\ }\textbf {\bibinfo {volume} {46}},\
  \bibinfo {pages} {67--97} (\bibinfo {year} {2018})}\BibitemShut {NoStop}%
\bibitem [{\citenamefont {Pan}\ \emph {et~al.}(2013)\citenamefont {Pan},
  \citenamefont {Spanu}, \citenamefont {Harrison}, \citenamefont {Sverjensky},\
  and\ \citenamefont {Galli}}]{pan_dielectric_2013}%
  \BibitemOpen
  \bibfield  {author} {\bibinfo {author} {\bibfnamefont {D.}~\bibnamefont
  {Pan}}, \bibinfo {author} {\bibfnamefont {L.}~\bibnamefont {Spanu}}, \bibinfo
  {author} {\bibfnamefont {B.}~\bibnamefont {Harrison}}, \bibinfo {author}
  {\bibfnamefont {D.~A.}\ \bibnamefont {Sverjensky}}, \ and\ \bibinfo {author}
  {\bibfnamefont {G.}~\bibnamefont {Galli}},\ }\bibfield  {title} {\enquote
  {\bibinfo {title} {Dielectric properties of water under extreme conditions
  and transport of carbonates in the deep {E}arth},}\ }\href {\doibase
  10.1073/pnas.1221581110} {\bibfield  {journal} {\bibinfo  {journal} {Proc
  Natl Acad Sci USA}\ }\textbf {\bibinfo {volume} {110}},\ \bibinfo {pages}
  {6646--6650} (\bibinfo {year} {2013})}\BibitemShut {NoStop}%
\bibitem [{\citenamefont {Neumann}\ and\ \citenamefont
  {Steinhauser}(1984)}]{neumann_computer_1984}%
  \BibitemOpen
  \bibfield  {author} {\bibinfo {author} {\bibfnamefont {M.}~\bibnamefont
  {Neumann}}\ and\ \bibinfo {author} {\bibfnamefont {O.}~\bibnamefont
  {Steinhauser}},\ }\bibfield  {title} {\enquote {\bibinfo {title} {Computer
  simulation and the dielectric constant of polarizable polar systems},}\
  }\href {\doibase 10.1016/0009-2614(84)85384-1} {\bibfield  {journal}
  {\bibinfo  {journal} {Chem Phys Lett}\ }\textbf {\bibinfo {volume} {106}},\
  \bibinfo {pages} {563--569} (\bibinfo {year} {1984})}\BibitemShut {NoStop}%
\bibitem [{\citenamefont {Gereben}\ and\ \citenamefont
  {Pusztai}(2011)}]{gereben_accurate_2011}%
  \BibitemOpen
  \bibfield  {author} {\bibinfo {author} {\bibfnamefont {Orsolya}\ \bibnamefont
  {Gereben}}\ and\ \bibinfo {author} {\bibfnamefont {L{\'a}szl{\'o}}\
  \bibnamefont {Pusztai}},\ }\bibfield  {title} {\enquote {\bibinfo {title} {On
  the accurate calculation of the dielectric constant from molecular dynamics
  simulations: The case of {SPC}/{E} and {SWM4}-{DP} water},}\ }\href {\doibase
  10.1016/j.cplett.2011.02.064} {\bibfield  {journal} {\bibinfo  {journal}
  {Chem Phys Lett}\ }\textbf {\bibinfo {volume} {507}},\ \bibinfo {pages}
  {80--83} (\bibinfo {year} {2011})}\BibitemShut {NoStop}%
\bibitem [{\citenamefont {Marzari}\ \emph {et~al.}(2012)\citenamefont
  {Marzari}, \citenamefont {Mostofi}, \citenamefont {Yates}, \citenamefont
  {Souza},\ and\ \citenamefont {Vanderbilt}}]{marzari_maximally_2012}%
  \BibitemOpen
  \bibfield  {author} {\bibinfo {author} {\bibfnamefont {Nicola}\ \bibnamefont
  {Marzari}}, \bibinfo {author} {\bibfnamefont {Arash~A.}\ \bibnamefont
  {Mostofi}}, \bibinfo {author} {\bibfnamefont {Jonathan~R.}\ \bibnamefont
  {Yates}}, \bibinfo {author} {\bibfnamefont {Ivo}\ \bibnamefont {Souza}}, \
  and\ \bibinfo {author} {\bibfnamefont {David}\ \bibnamefont {Vanderbilt}},\
  }\bibfield  {title} {\enquote {\bibinfo {title} {Maximally localized wannier
  functions: Theory and applications},}\ }\href {\doibase
  10.1103/RevModPhys.84.1419} {\bibfield  {journal} {\bibinfo  {journal} {Rev
  Mod Phys}\ }\textbf {\bibinfo {volume} {84}},\ \bibinfo {pages} {1419--1475}
  (\bibinfo {year} {2012})}\BibitemShut {NoStop}%
\bibitem [{\citenamefont {Behler}\ and\ \citenamefont
  {Parrinello}(2007)}]{behler_generalized_2007}%
  \BibitemOpen
  \bibfield  {author} {\bibinfo {author} {\bibfnamefont {Jörg}\ \bibnamefont
  {Behler}}\ and\ \bibinfo {author} {\bibfnamefont {Michele}\ \bibnamefont
  {Parrinello}},\ }\bibfield  {title} {\enquote {\bibinfo {title} {Generalized
  neural-network representation of high-dimensional potential-energy
  surfaces},}\ }\href {\doibase 10.1103/PhysRevLett.98.146401} {\bibfield
  {journal} {\bibinfo  {journal} {Phys Rev Lett}\ }\textbf {\bibinfo {volume}
  {98}},\ \bibinfo {pages} {146401} (\bibinfo {year} {2007})}\BibitemShut
  {NoStop}%
\bibitem [{\citenamefont {Bartók}\ \emph {et~al.}(2010)\citenamefont
  {Bartók}, \citenamefont {Payne}, \citenamefont {Kondor},\ and\ \citenamefont
  {Csányi}}]{bartok_gaussian_2010}%
  \BibitemOpen
  \bibfield  {author} {\bibinfo {author} {\bibfnamefont {Albert~P.}\
  \bibnamefont {Bartók}}, \bibinfo {author} {\bibfnamefont {Mike~C.}\
  \bibnamefont {Payne}}, \bibinfo {author} {\bibfnamefont {Risi}\ \bibnamefont
  {Kondor}}, \ and\ \bibinfo {author} {\bibfnamefont {Gábor}\ \bibnamefont
  {Csányi}},\ }\bibfield  {title} {\enquote {\bibinfo {title} {Gaussian
  approximation potentials: The accuracy of quantum mechanics, without the
  electrons},}\ }\href {\doibase 10.1103/PhysRevLett.104.136403} {\bibfield
  {journal} {\bibinfo  {journal} {Phys Rev Lett}\ }\textbf {\bibinfo {volume}
  {104}},\ \bibinfo {pages} {136403} (\bibinfo {year} {2010})}\BibitemShut
  {NoStop}%
\bibitem [{\citenamefont {Rupp}\ \emph {et~al.}(2012)\citenamefont {Rupp},
  \citenamefont {Tkatchenko}, \citenamefont {Müller},\ and\ \citenamefont {von
  Lilienfeld}}]{rupp_fast_2012}%
  \BibitemOpen
  \bibfield  {author} {\bibinfo {author} {\bibfnamefont {Matthias}\
  \bibnamefont {Rupp}}, \bibinfo {author} {\bibfnamefont {Alexandre}\
  \bibnamefont {Tkatchenko}}, \bibinfo {author} {\bibfnamefont {Klaus-Robert}\
  \bibnamefont {Müller}}, \ and\ \bibinfo {author} {\bibfnamefont
  {O.~Anatole}\ \bibnamefont {von Lilienfeld}},\ }\bibfield  {title} {\enquote
  {\bibinfo {title} {Fast and accurate modeling of molecular atomization
  energies with machine learning},}\ }\href {\doibase
  10.1103/PhysRevLett.108.058301} {\bibfield  {journal} {\bibinfo  {journal}
  {Phys Rev Lett}\ }\textbf {\bibinfo {volume} {108}},\ \bibinfo {pages}
  {058301} (\bibinfo {year} {2012})}\BibitemShut {NoStop}%
\bibitem [{\citenamefont {Smith}\ \emph {et~al.}(2017)\citenamefont {Smith},
  \citenamefont {Isayev},\ and\ \citenamefont {Roitberg}}]{smith_ani-1_2017}%
  \BibitemOpen
  \bibfield  {author} {\bibinfo {author} {\bibfnamefont {J.~S.}\ \bibnamefont
  {Smith}}, \bibinfo {author} {\bibfnamefont {O.}~\bibnamefont {Isayev}}, \
  and\ \bibinfo {author} {\bibfnamefont {A.~E.}\ \bibnamefont {Roitberg}},\
  }\bibfield  {title} {\enquote {\bibinfo {title} {{ANI}-1: an extensible
  neural network potential with {DFT} accuracy at force field computational
  cost},}\ }\href {\doibase 10.1039/C6SC05720A} {\bibfield  {journal} {\bibinfo
   {journal} {Chem Sci}\ }\textbf {\bibinfo {volume} {8}},\ \bibinfo {pages}
  {3192--3203} (\bibinfo {year} {2017})}\BibitemShut {NoStop}%
\bibitem [{\citenamefont {Zhang}\ \emph {et~al.}(2018)\citenamefont {Zhang},
  \citenamefont {Han}, \citenamefont {Wang}, \citenamefont {Car},\ and\
  \citenamefont {E}}]{zhang_deep_2018}%
  \BibitemOpen
  \bibfield  {author} {\bibinfo {author} {\bibfnamefont {Linfeng}\ \bibnamefont
  {Zhang}}, \bibinfo {author} {\bibfnamefont {Jiequn}\ \bibnamefont {Han}},
  \bibinfo {author} {\bibfnamefont {Han}\ \bibnamefont {Wang}}, \bibinfo
  {author} {\bibfnamefont {Roberto}\ \bibnamefont {Car}}, \ and\ \bibinfo
  {author} {\bibfnamefont {Weinan}\ \bibnamefont {E}},\ }\bibfield  {title}
  {\enquote {\bibinfo {title} {Deep potential molecular dynamics: A scalable
  model with the accuracy of quantum mechanics},}\ }\href {\doibase
  10.1103/PhysRevLett.120.143001} {\bibfield  {journal} {\bibinfo  {journal}
  {Phys Rev Lett}\ }\textbf {\bibinfo {volume} {120}},\ \bibinfo {pages}
  {143001} (\bibinfo {year} {2018})}\BibitemShut {NoStop}%
\bibitem [{\citenamefont {Umeda}\ \emph {et~al.}(2019)\citenamefont {Umeda},
  \citenamefont {Hayashi}, \citenamefont {Moriwake},\ and\ \citenamefont
  {Tanaka}}]{umeda_prediction_2019}%
  \BibitemOpen
  \bibfield  {author} {\bibinfo {author} {\bibfnamefont {Yuji}\ \bibnamefont
  {Umeda}}, \bibinfo {author} {\bibfnamefont {Hiroyuki}\ \bibnamefont
  {Hayashi}}, \bibinfo {author} {\bibfnamefont {Hiroki}\ \bibnamefont
  {Moriwake}}, \ and\ \bibinfo {author} {\bibfnamefont {Isao}\ \bibnamefont
  {Tanaka}},\ }\bibfield  {title} {\enquote {\bibinfo {title} {Prediction of
  dielectric constants using a combination of first principles calculations and
  machine learning},}\ }\href {\doibase 10.7567/1347-4065/ab34d6} {\bibfield
  {journal} {\bibinfo  {journal} {Jpn J Appl Phys}\ }\textbf {\bibinfo {volume}
  {58}},\ \bibinfo {pages} {SLLC01} (\bibinfo {year} {2019})}\BibitemShut
  {NoStop}%
\bibitem [{\citenamefont {Sharma}\ \emph {et~al.}(2007)\citenamefont {Sharma},
  \citenamefont {Resta},\ and\ \citenamefont {Car}}]{sharma_dipolar_2007}%
  \BibitemOpen
  \bibfield  {author} {\bibinfo {author} {\bibfnamefont {Manu}\ \bibnamefont
  {Sharma}}, \bibinfo {author} {\bibfnamefont {Raffaele}\ \bibnamefont
  {Resta}}, \ and\ \bibinfo {author} {\bibfnamefont {Roberto}\ \bibnamefont
  {Car}},\ }\bibfield  {title} {\enquote {\bibinfo {title} {Dipolar
  correlations and the dielectric permittivity of water},}\ }\href {\doibase
  10.1103/PhysRevLett.98.247401} {\bibfield  {journal} {\bibinfo  {journal}
  {Phys Rev Lett}\ }\textbf {\bibinfo {volume} {98}},\ \bibinfo {pages}
  {247401} (\bibinfo {year} {2007})}\BibitemShut {NoStop}%
\bibitem [{\citenamefont {Baroni}\ \emph {et~al.}(2001)\citenamefont {Baroni},
  \citenamefont {De~Gironcoli}, \citenamefont {Dal~Corso},\ and\ \citenamefont
  {Giannozzi}}]{baroni2001phonons}%
  \BibitemOpen
  \bibfield  {author} {\bibinfo {author} {\bibfnamefont {Stefano}\ \bibnamefont
  {Baroni}}, \bibinfo {author} {\bibfnamefont {Stefano}\ \bibnamefont
  {De~Gironcoli}}, \bibinfo {author} {\bibfnamefont {Andrea}\ \bibnamefont
  {Dal~Corso}}, \ and\ \bibinfo {author} {\bibfnamefont {Paolo}\ \bibnamefont
  {Giannozzi}},\ }\bibfield  {title} {\enquote {\bibinfo {title} {Phonons and
  related crystal properties from density-functional perturbation theory},}\
  }\href@noop {} {\bibfield  {journal} {\bibinfo  {journal} {Rev Mod Phys}\
  }\textbf {\bibinfo {volume} {73}},\ \bibinfo {pages} {515} (\bibinfo {year}
  {2001})}\BibitemShut {NoStop}%
\bibitem [{\citenamefont {Pan}\ \emph {et~al.}(2014)\citenamefont {Pan},
  \citenamefont {Wan},\ and\ \citenamefont {Galli}}]{pan2014refractive}%
  \BibitemOpen
  \bibfield  {author} {\bibinfo {author} {\bibfnamefont {Ding}\ \bibnamefont
  {Pan}}, \bibinfo {author} {\bibfnamefont {Quan}\ \bibnamefont {Wan}}, \ and\
  \bibinfo {author} {\bibfnamefont {Giulia}\ \bibnamefont {Galli}},\ }\bibfield
   {title} {\enquote {\bibinfo {title} {The refractive index and electronic gap
  of water and ice increase with increasing pressure},}\ }\href {\doibase
  10.1038/ncomms4919} {\bibfield  {journal} {\bibinfo  {journal} {Nat Commun}\
  }\textbf {\bibinfo {volume} {5}},\ \bibinfo {pages} {3919} (\bibinfo {year}
  {2014})}\BibitemShut {NoStop}%
\bibitem [{\citenamefont {Grisafi}\ \emph {et~al.}(2018)\citenamefont
  {Grisafi}, \citenamefont {Wilkins}, \citenamefont {Cs\'anyi},\ and\
  \citenamefont {Ceriotti}}]{PhysRevLett.120.036002}%
  \BibitemOpen
  \bibfield  {author} {\bibinfo {author} {\bibfnamefont {Andrea}\ \bibnamefont
  {Grisafi}}, \bibinfo {author} {\bibfnamefont {David~M.}\ \bibnamefont
  {Wilkins}}, \bibinfo {author} {\bibfnamefont {G\'abor}\ \bibnamefont
  {Cs\'anyi}}, \ and\ \bibinfo {author} {\bibfnamefont {Michele}\ \bibnamefont
  {Ceriotti}},\ }\bibfield  {title} {\enquote {\bibinfo {title}
  {Symmetry-adapted machine learning for tensorial properties of atomistic
  systems},}\ }\href {\doibase 10.1103/PhysRevLett.120.036002} {\bibfield
  {journal} {\bibinfo  {journal} {Phys Rev Lett}\ }\textbf {\bibinfo {volume}
  {120}},\ \bibinfo {pages} {036002} (\bibinfo {year} {2018})}\BibitemShut
  {NoStop}%
\bibitem [{\citenamefont {Abadi}\ \emph {et~al.}(2015)\citenamefont {Abadi},
  \citenamefont {Agarwal}, \citenamefont {Barham}, \citenamefont {Brevdo},
  \citenamefont {Chen}, \citenamefont {Citro}, \citenamefont {Corrado},
  \citenamefont {Davis}, \citenamefont {Dean}, \citenamefont {Devin},
  \citenamefont {Ghemawat}, \citenamefont {Goodfellow}, \citenamefont {Harp},
  \citenamefont {Irving}, \citenamefont {Isard}, \citenamefont {Jia},
  \citenamefont {Jozefowicz}, \citenamefont {Kaiser}, \citenamefont {Kudlur},
  \citenamefont {Levenberg}, \citenamefont {Man\'{e}}, \citenamefont {Monga},
  \citenamefont {Moore}, \citenamefont {Murray}, \citenamefont {Olah},
  \citenamefont {Schuster}, \citenamefont {Shlens}, \citenamefont {Steiner},
  \citenamefont {Sutskever}, \citenamefont {Talwar}, \citenamefont {Tucker},
  \citenamefont {Vanhoucke}, \citenamefont {Vasudevan}, \citenamefont
  {Vi\'{e}gas}, \citenamefont {Vinyals}, \citenamefont {Warden}, \citenamefont
  {Wattenberg}, \citenamefont {Wicke}, \citenamefont {Yu},\ and\ \citenamefont
  {Zheng}}]{tensorflow2015-whitepaper}%
  \BibitemOpen
  \bibfield  {author} {\bibinfo {author} {\bibfnamefont {Mart\'{\i}n}\
  \bibnamefont {Abadi}}, \bibinfo {author} {\bibfnamefont {Ashish}\
  \bibnamefont {Agarwal}}, \bibinfo {author} {\bibfnamefont {Paul}\
  \bibnamefont {Barham}}, \bibinfo {author} {\bibfnamefont {Eugene}\
  \bibnamefont {Brevdo}}, \bibinfo {author} {\bibfnamefont {Zhifeng}\
  \bibnamefont {Chen}}, \bibinfo {author} {\bibfnamefont {Craig}\ \bibnamefont
  {Citro}}, \bibinfo {author} {\bibfnamefont {Greg~S.}\ \bibnamefont
  {Corrado}}, \bibinfo {author} {\bibfnamefont {Andy}\ \bibnamefont {Davis}},
  \bibinfo {author} {\bibfnamefont {Jeffrey}\ \bibnamefont {Dean}}, \bibinfo
  {author} {\bibfnamefont {Matthieu}\ \bibnamefont {Devin}}, \bibinfo {author}
  {\bibfnamefont {Sanjay}\ \bibnamefont {Ghemawat}}, \bibinfo {author}
  {\bibfnamefont {Ian}\ \bibnamefont {Goodfellow}}, \bibinfo {author}
  {\bibfnamefont {Andrew}\ \bibnamefont {Harp}}, \bibinfo {author}
  {\bibfnamefont {Geoffrey}\ \bibnamefont {Irving}}, \bibinfo {author}
  {\bibfnamefont {Michael}\ \bibnamefont {Isard}}, \bibinfo {author}
  {\bibfnamefont {Yangqing}\ \bibnamefont {Jia}}, \bibinfo {author}
  {\bibfnamefont {Rafal}\ \bibnamefont {Jozefowicz}}, \bibinfo {author}
  {\bibfnamefont {Lukasz}\ \bibnamefont {Kaiser}}, \bibinfo {author}
  {\bibfnamefont {Manjunath}\ \bibnamefont {Kudlur}}, \bibinfo {author}
  {\bibfnamefont {Josh}\ \bibnamefont {Levenberg}}, \bibinfo {author}
  {\bibfnamefont {Dandelion}\ \bibnamefont {Man\'{e}}}, \bibinfo {author}
  {\bibfnamefont {Rajat}\ \bibnamefont {Monga}}, \bibinfo {author}
  {\bibfnamefont {Sherry}\ \bibnamefont {Moore}}, \bibinfo {author}
  {\bibfnamefont {Derek}\ \bibnamefont {Murray}}, \bibinfo {author}
  {\bibfnamefont {Chris}\ \bibnamefont {Olah}}, \bibinfo {author}
  {\bibfnamefont {Mike}\ \bibnamefont {Schuster}}, \bibinfo {author}
  {\bibfnamefont {Jonathon}\ \bibnamefont {Shlens}}, \bibinfo {author}
  {\bibfnamefont {Benoit}\ \bibnamefont {Steiner}}, \bibinfo {author}
  {\bibfnamefont {Ilya}\ \bibnamefont {Sutskever}}, \bibinfo {author}
  {\bibfnamefont {Kunal}\ \bibnamefont {Talwar}}, \bibinfo {author}
  {\bibfnamefont {Paul}\ \bibnamefont {Tucker}}, \bibinfo {author}
  {\bibfnamefont {Vincent}\ \bibnamefont {Vanhoucke}}, \bibinfo {author}
  {\bibfnamefont {Vijay}\ \bibnamefont {Vasudevan}}, \bibinfo {author}
  {\bibfnamefont {Fernanda}\ \bibnamefont {Vi\'{e}gas}}, \bibinfo {author}
  {\bibfnamefont {Oriol}\ \bibnamefont {Vinyals}}, \bibinfo {author}
  {\bibfnamefont {Pete}\ \bibnamefont {Warden}}, \bibinfo {author}
  {\bibfnamefont {Martin}\ \bibnamefont {Wattenberg}}, \bibinfo {author}
  {\bibfnamefont {Martin}\ \bibnamefont {Wicke}}, \bibinfo {author}
  {\bibfnamefont {Yuan}\ \bibnamefont {Yu}}, \ and\ \bibinfo {author}
  {\bibfnamefont {Xiaoqiang}\ \bibnamefont {Zheng}},\ }\href
  {https://www.tensorflow.org/} {\enquote {\bibinfo {title} {{TensorFlow}:
  Large-scale machine learning on heterogeneous systems},}\ } (\bibinfo {year}
  {2015}),\ \bibinfo {note} {software available from
  tensorflow.org}\BibitemShut {NoStop}%
\bibitem [{\citenamefont {Kingma}\ and\ \citenamefont
  {Ba}(2014)}]{kingma_adam_2017}%
  \BibitemOpen
  \bibfield  {author} {\bibinfo {author} {\bibfnamefont {Diederik~P.}\
  \bibnamefont {Kingma}}\ and\ \bibinfo {author} {\bibfnamefont {Jimmy}\
  \bibnamefont {Ba}},\ }\bibfield  {title} {\enquote {\bibinfo {title} {Adam: A
  method for stochastic optimization},}\ }\href
  {http://arxiv.org/abs/1412.6980} {\bibfield  {journal} {\bibinfo  {journal}
  {arXiv:1412.6980}\ } (\bibinfo {year} {2014})}\BibitemShut {NoStop}%
\bibitem [{\citenamefont {Ng}(2004)}]{10.1145/1015330.1015435}%
  \BibitemOpen
  \bibfield  {author} {\bibinfo {author} {\bibfnamefont {Andrew~Y.}\
  \bibnamefont {Ng}},\ }\bibfield  {title} {\enquote {\bibinfo {title} {Feature
  selection, {L}1 vs. {L}2 regularization, and rotational invariance},}\ }in\
  \href {\doibase 10.1145/1015330.1015435} {\emph {\bibinfo {booktitle}
  {Proceedings of the Twenty-First International Conference on Machine
  Learning}}},\ \bibinfo {series and number} {ICML ’04}\ (\bibinfo
  {publisher} {Association for Computing Machinery},\ \bibinfo {address} {New
  York, NY, USA},\ \bibinfo {year} {2004})\ p.~\bibinfo {pages}
  {78}\BibitemShut {NoStop}%
\bibitem [{\citenamefont {Perdew}\ \emph {et~al.}(1996)\citenamefont {Perdew},
  \citenamefont {Burke},\ and\ \citenamefont {Ernzerhof}}]{Perdew1996}%
  \BibitemOpen
  \bibfield  {author} {\bibinfo {author} {\bibfnamefont {John~P.}\ \bibnamefont
  {Perdew}}, \bibinfo {author} {\bibfnamefont {Kieron}\ \bibnamefont {Burke}},
  \ and\ \bibinfo {author} {\bibfnamefont {Matthias}\ \bibnamefont
  {Ernzerhof}},\ }\bibfield  {title} {\enquote {\bibinfo {title} {Generalized
  gradient approximation made simple},}\ }\href {\doibase
  10.1103/PhysRevLett.77.3865} {\bibfield  {journal} {\bibinfo  {journal} {Phys
  Rev Lett}\ }\textbf {\bibinfo {volume} {77}},\ \bibinfo {pages} {3865--3868}
  (\bibinfo {year} {1996})}\BibitemShut {NoStop}%
\bibitem [{\citenamefont {Zhang}\ \emph {et~al.}(2011)\citenamefont {Zhang},
  \citenamefont {Donadio}, \citenamefont {Gygi},\ and\ \citenamefont
  {Galli}}]{zhang2011first}%
  \BibitemOpen
  \bibfield  {author} {\bibinfo {author} {\bibfnamefont {Cui}\ \bibnamefont
  {Zhang}}, \bibinfo {author} {\bibfnamefont {Davide}\ \bibnamefont {Donadio}},
  \bibinfo {author} {\bibfnamefont {Francois}\ \bibnamefont {Gygi}}, \ and\
  \bibinfo {author} {\bibfnamefont {Giulia}\ \bibnamefont {Galli}},\ }\bibfield
   {title} {\enquote {\bibinfo {title} {First principles simulations of the
  infrared spectrum of liquid water using hybrid density functionals},}\
  }\href@noop {} {\bibfield  {journal} {\bibinfo  {journal} {J Chem Theory
  Comput}\ }\textbf {\bibinfo {volume} {7}},\ \bibinfo {pages} {1443--1449}
  (\bibinfo {year} {2011})}\BibitemShut {NoStop}%
\bibitem [{\citenamefont {Pan}\ and\ \citenamefont {Galli}(2016)}]{Pan2016}%
  \BibitemOpen
  \bibfield  {author} {\bibinfo {author} {\bibfnamefont {D.}~\bibnamefont
  {Pan}}\ and\ \bibinfo {author} {\bibfnamefont {G.}~\bibnamefont {Galli}},\
  }\bibfield  {title} {\enquote {\bibinfo {title} {The fate of carbon dioxide
  in water-rich fluids under extreme conditions},}\ }\href {\doibase
  10.1126/sciadv.1601278} {\bibfield  {journal} {\bibinfo  {journal} {Sci Adv}\
  }\textbf {\bibinfo {volume} {2}},\ \bibinfo {pages} {e1601278} (\bibinfo
  {year} {2016})}\BibitemShut {NoStop}%
\bibitem [{\citenamefont {Berendsen}\ \emph {et~al.}(1987)\citenamefont
  {Berendsen}, \citenamefont {Grigera},\ and\ \citenamefont
  {Straatsma}}]{berendsen_missing_1987}%
  \BibitemOpen
  \bibfield  {author} {\bibinfo {author} {\bibfnamefont {H.~J.~C.}\
  \bibnamefont {Berendsen}}, \bibinfo {author} {\bibfnamefont {J.~R.}\
  \bibnamefont {Grigera}}, \ and\ \bibinfo {author} {\bibfnamefont {T.~P.}\
  \bibnamefont {Straatsma}},\ }\bibfield  {title} {\enquote {\bibinfo {title}
  {The missing term in effective pair potentials},}\ }\href {\doibase
  10.1021/j100308a038} {\bibfield  {journal} {\bibinfo  {journal} {J Phys
  Chem}\ }\textbf {\bibinfo {volume} {91}},\ \bibinfo {pages} {6269--6271}
  (\bibinfo {year} {1987})}\BibitemShut {NoStop}%
\bibitem [{\citenamefont {Abascal}\ and\ \citenamefont
  {Vega}(2005)}]{doi:10.1063/1.2121687}%
  \BibitemOpen
  \bibfield  {author} {\bibinfo {author} {\bibfnamefont {J.~L.~F.}\
  \bibnamefont {Abascal}}\ and\ \bibinfo {author} {\bibfnamefont
  {C.}~\bibnamefont {Vega}},\ }\bibfield  {title} {\enquote {\bibinfo {title}
  {A general purpose model for the condensed phases of water: {TIP4P}/2005},}\
  }\href {\doibase 10.1063/1.2121687} {\bibfield  {journal} {\bibinfo
  {journal} {J Chem Phys}\ }\textbf {\bibinfo {volume} {123}},\ \bibinfo
  {pages} {234505} (\bibinfo {year} {2005})}\BibitemShut {NoStop}%
\bibitem [{\citenamefont {Zhang}\ and\ \citenamefont
  {Duan}(2005)}]{zhang_prediction_2005}%
  \BibitemOpen
  \bibfield  {author} {\bibinfo {author} {\bibfnamefont {Zhigang}\ \bibnamefont
  {Zhang}}\ and\ \bibinfo {author} {\bibfnamefont {Zhenhao}\ \bibnamefont
  {Duan}},\ }\bibfield  {title} {\enquote {\bibinfo {title} {Prediction of the
  {PVT} properties of water over wide range of temperatures and pressures from
  molecular dynamics simulation},}\ }\href {\doibase
  https://doi.org/10.1016/j.pepi.2004.11.003} {\bibfield  {journal} {\bibinfo
  {journal} {Phys Earth Planet Inter}\ }\textbf {\bibinfo {volume} {149}},\
  \bibinfo {pages} {335 -- 354} (\bibinfo {year} {2005})}\BibitemShut {NoStop}%
\bibitem [{\citenamefont {Kirkwood}(1939)}]{kirkwood1939dielectric}%
  \BibitemOpen
  \bibfield  {author} {\bibinfo {author} {\bibfnamefont {John~G}\ \bibnamefont
  {Kirkwood}},\ }\bibfield  {title} {\enquote {\bibinfo {title} {The dielectric
  polarization of polar liquids},}\ }\href@noop {} {\bibfield  {journal}
  {\bibinfo  {journal} {J Chem Phys}\ }\textbf {\bibinfo {volume} {7}},\
  \bibinfo {pages} {911--919} (\bibinfo {year} {1939})}\BibitemShut {NoStop}%
\bibitem [{\citenamefont {Aragones}\ \emph {et~al.}(2011)\citenamefont
  {Aragones}, \citenamefont {MacDowell},\ and\ \citenamefont
  {Vega}}]{aragones2011dielectric}%
  \BibitemOpen
  \bibfield  {author} {\bibinfo {author} {\bibfnamefont {JL}~\bibnamefont
  {Aragones}}, \bibinfo {author} {\bibfnamefont {LG}~\bibnamefont {MacDowell}},
  \ and\ \bibinfo {author} {\bibfnamefont {C}~\bibnamefont {Vega}},\ }\bibfield
   {title} {\enquote {\bibinfo {title} {Dielectric constant of ices and water:
  a lesson about water interactions},}\ }\href@noop {} {\bibfield  {journal}
  {\bibinfo  {journal} {J Phys Chem A}\ }\textbf {\bibinfo {volume} {115}},\
  \bibinfo {pages} {5745--5758} (\bibinfo {year} {2011})}\BibitemShut {NoStop}%
\bibitem [{\citenamefont {Zhuang}\ \emph {et~al.}(2020)\citenamefont {Zhuang},
  \citenamefont {Ye}, \citenamefont {Pan},\ and\ \citenamefont
  {Li}}]{LinZhuang-43101}%
  \BibitemOpen
  \bibfield  {author} {\bibinfo {author} {\bibfnamefont {Lin}\ \bibnamefont
  {Zhuang}}, \bibinfo {author} {\bibfnamefont {Qijun}\ \bibnamefont {Ye}},
  \bibinfo {author} {\bibfnamefont {Ding}\ \bibnamefont {Pan}}, \ and\ \bibinfo
  {author} {\bibfnamefont {Xin-Zheng}\ \bibnamefont {Li}},\ }\bibfield  {title}
  {\enquote {\bibinfo {title} {Discriminating high-pressure water phases using
  rare-event determined ionic dynamical properties},}\ }\href {\doibase
  10.1088/0256-307X/37/4/043101} {\bibfield  {journal} {\bibinfo  {journal}
  {Chinese Phys Lett}\ }\textbf {\bibinfo {volume} {37}},\ \bibinfo {eid}
  {043101} (\bibinfo {year} {2020})}\BibitemShut {NoStop}%
\bibitem [{\citenamefont {Schwegler}\ \emph {et~al.}(2000)\citenamefont
  {Schwegler}, \citenamefont {Galli},\ and\ \citenamefont
  {Gygi}}]{schwegler2000water}%
  \BibitemOpen
  \bibfield  {author} {\bibinfo {author} {\bibfnamefont {Eric}\ \bibnamefont
  {Schwegler}}, \bibinfo {author} {\bibfnamefont {Giulia}\ \bibnamefont
  {Galli}}, \ and\ \bibinfo {author} {\bibfnamefont {Fran{\c{c}}ois}\
  \bibnamefont {Gygi}},\ }\bibfield  {title} {\enquote {\bibinfo {title} {Water
  under pressure},}\ }\href@noop {} {\bibfield  {journal} {\bibinfo  {journal}
  {Physical Review Letters}\ }\textbf {\bibinfo {volume} {84}},\ \bibinfo
  {pages} {2429} (\bibinfo {year} {2000})}\BibitemShut {NoStop}%
\bibitem [{\citenamefont {van~der Spoel}\ \emph {et~al.}(1998)\citenamefont
  {van~der Spoel}, \citenamefont {van Maaren},\ and\ \citenamefont
  {Berendsen}}]{van_der_spoel_systematic_1998}%
  \BibitemOpen
  \bibfield  {author} {\bibinfo {author} {\bibfnamefont {David}\ \bibnamefont
  {van~der Spoel}}, \bibinfo {author} {\bibfnamefont {Paul~J.}\ \bibnamefont
  {van Maaren}}, \ and\ \bibinfo {author} {\bibfnamefont {Herman J.~C.}\
  \bibnamefont {Berendsen}},\ }\bibfield  {title} {\enquote {\bibinfo {title}
  {A systematic study of water models for molecular simulation: {Derivation} of
  water models optimized for use with a reaction field},}\ }\href {\doibase
  10.1063/1.476482} {\bibfield  {journal} {\bibinfo  {journal} {J Chem Phys}\
  }\textbf {\bibinfo {volume} {108}},\ \bibinfo {pages} {10220--10230}
  (\bibinfo {year} {1998})}\BibitemShut {NoStop}%
\bibitem [{\citenamefont {Elton}(2017)}]{elton_origin_2017}%
  \BibitemOpen
  \bibfield  {author} {\bibinfo {author} {\bibfnamefont {Daniel~C.}\
  \bibnamefont {Elton}},\ }\bibfield  {title} {\enquote {\bibinfo {title} {The
  origin of the {Debye} relaxation in liquid water and fitting the high
  frequency excess response},}\ }\href {\doibase 10.1039/C7CP02884A} {\bibfield
   {journal} {\bibinfo  {journal} {Phys Chem Chem Phys}\ }\textbf {\bibinfo
  {volume} {19}},\ \bibinfo {pages} {18739--18749} (\bibinfo {year}
  {2017})}\BibitemShut {NoStop}%
\bibitem [{\citenamefont {Zhang}\ \emph {et~al.}(2020)\citenamefont {Zhang},
  \citenamefont {Chen}, \citenamefont {Wu}, \citenamefont {Wang}, \citenamefont
  {E},\ and\ \citenamefont {Car}}]{zhang2019weinan}%
  \BibitemOpen
  \bibfield  {author} {\bibinfo {author} {\bibfnamefont {Linfeng}\ \bibnamefont
  {Zhang}}, \bibinfo {author} {\bibfnamefont {Mohan}\ \bibnamefont {Chen}},
  \bibinfo {author} {\bibfnamefont {Xifan}\ \bibnamefont {Wu}}, \bibinfo
  {author} {\bibfnamefont {Han}\ \bibnamefont {Wang}}, \bibinfo {author}
  {\bibfnamefont {Weinan}\ \bibnamefont {E}}, \ and\ \bibinfo {author}
  {\bibfnamefont {Roberto}\ \bibnamefont {Car}},\ }\bibfield  {title} {\enquote
  {\bibinfo {title} {Deep neural network for the dielectric response of
  insulators},}\ }\href {\doibase 10.1103/physrevb.102.041121} {\bibfield
  {journal} {\bibinfo  {journal} {Physical Review B}\ }\textbf {\bibinfo
  {volume} {102}} (\bibinfo {year} {2020}),\
  10.1103/physrevb.102.041121}\BibitemShut {NoStop}%
\bibitem [{\citenamefont {Hazen}\ and\ \citenamefont
  {Schiffries}(2013)}]{hazen2013deep}%
  \BibitemOpen
  \bibfield  {author} {\bibinfo {author} {\bibfnamefont {Robert~M}\
  \bibnamefont {Hazen}}\ and\ \bibinfo {author} {\bibfnamefont {Craig~M}\
  \bibnamefont {Schiffries}},\ }\bibfield  {title} {\enquote {\bibinfo {title}
  {Why deep carbon?}}\ }\href@noop {} {\bibfield  {journal} {\bibinfo
  {journal} {Rev Mineral Geochem}\ }\textbf {\bibinfo {volume} {75}},\ \bibinfo
  {pages} {1--6} (\bibinfo {year} {2013})}\BibitemShut {NoStop}%
\bibitem [{\citenamefont {Huang}\ and\ \citenamefont
  {Sverjensky}(2019)}]{huang2019extended}%
  \BibitemOpen
  \bibfield  {author} {\bibinfo {author} {\bibfnamefont {Fang}\ \bibnamefont
  {Huang}}\ and\ \bibinfo {author} {\bibfnamefont {Dimitri~A}\ \bibnamefont
  {Sverjensky}},\ }\bibfield  {title} {\enquote {\bibinfo {title} {Extended
  deep earth water model for predicting major element mantle metasomatism},}\
  }\href@noop {} {\bibfield  {journal} {\bibinfo  {journal} {Geochimica et
  Cosmochimica Acta}\ }\textbf {\bibinfo {volume} {254}},\ \bibinfo {pages}
  {192--230} (\bibinfo {year} {2019})}\BibitemShut {NoStop}%
\bibitem [{\citenamefont {Flyvbjerg}\ and\ \citenamefont
  {Petersen}(1989)}]{flyvbjerg1989error}%
  \BibitemOpen
  \bibfield  {author} {\bibinfo {author} {\bibfnamefont {Henrik}\ \bibnamefont
  {Flyvbjerg}}\ and\ \bibinfo {author} {\bibfnamefont {Henrik~Gordon}\
  \bibnamefont {Petersen}},\ }\bibfield  {title} {\enquote {\bibinfo {title}
  {Error estimates on averages of correlated data},}\ }\href@noop {} {\bibfield
   {journal} {\bibinfo  {journal} {J Chem Phys}\ }\textbf {\bibinfo {volume}
  {91}},\ \bibinfo {pages} {461--466} (\bibinfo {year} {1989})}\BibitemShut
  {NoStop}%
\end{thebibliography}%


\begin{thebibliography}{18}%
\makeatletter
\providecommand \@ifxundefined [1]{%
 \@ifx{#1\undefined}
}%
\providecommand \@ifnum [1]{%
 \ifnum #1\expandafter \@firstoftwo
 \else \expandafter \@secondoftwo
 \fi
}%
\providecommand \@ifx [1]{%
 \ifx #1\expandafter \@firstoftwo
 \else \expandafter \@secondoftwo
 \fi
}%
\providecommand \natexlab [1]{#1}%
\providecommand \enquote  [1]{``#1''}%
\providecommand \bibnamefont  [1]{#1}%
\providecommand \bibfnamefont [1]{#1}%
\providecommand \citenamefont [1]{#1}%
\providecommand \href@noop [0]{\@secondoftwo}%
\providecommand \href [0]{\begingroup \@sanitize@url \@href}%
\providecommand \@href[1]{\@@startlink{#1}\@@href}%
\providecommand \@@href[1]{\endgroup#1\@@endlink}%
\providecommand \@sanitize@url [0]{\catcode `\\12\catcode `\$12\catcode
  `\&12\catcode `\#12\catcode `\^12\catcode `\_12\catcode `\%12\relax}%
\providecommand \@@startlink[1]{}%
\providecommand \@@endlink[0]{}%
\providecommand \url  [0]{\begingroup\@sanitize@url \@url }%
\providecommand \@url [1]{\endgroup\@href {#1}{\urlprefix }}%
\providecommand \urlprefix  [0]{URL }%
\providecommand \Eprint [0]{\href }%
\providecommand \doibase [0]{http://dx.doi.org/}%
\providecommand \selectlanguage [0]{\@gobble}%
\providecommand \bibinfo  [0]{\@secondoftwo}%
\providecommand \bibfield  [0]{\@secondoftwo}%
\providecommand \translation [1]{[#1]}%
\providecommand \BibitemOpen [0]{}%
\providecommand \bibitemStop [0]{}%
\providecommand \bibitemNoStop [0]{.\EOS\space}%
\providecommand \EOS [0]{\spacefactor3000\relax}%
\providecommand \BibitemShut  [1]{\csname bibitem#1\endcsname}%
\let\auto@bib@innerbib\@empty
%</preamble>
\bibitem [{\citenamefont {Gygi}(2008)}]{gygi_architecture_2008}%
  \BibitemOpen
  \bibfield  {author} {\bibinfo {author} {\bibfnamefont {F.}~\bibnamefont
  {Gygi}},\ }\href {\doibase 10.1147/rd.521.0137} {\bibfield  {journal}
  {\bibinfo  {journal} {IBM J Res Develop}\ }\textbf {\bibinfo {volume} {52}},\
  \bibinfo {pages} {137} (\bibinfo {year} {2008})}\BibitemShut {NoStop}%
\bibitem [{\citenamefont {Perdew}\ \emph {et~al.}(1996)\citenamefont {Perdew},
  \citenamefont {Burke},\ and\ \citenamefont {Ernzerhof}}]{Perdew1996}%
  \BibitemOpen
  \bibfield  {author} {\bibinfo {author} {\bibfnamefont {J.~P.}\ \bibnamefont
  {Perdew}}, \bibinfo {author} {\bibfnamefont {K.}~\bibnamefont {Burke}}, \
  and\ \bibinfo {author} {\bibfnamefont {M.}~\bibnamefont {Ernzerhof}},\ }\href
  {\doibase 10.1103/PhysRevLett.77.3865} {\bibfield  {journal} {\bibinfo
  {journal} {Phys Rev Lett}\ }\textbf {\bibinfo {volume} {77}},\ \bibinfo
  {pages} {3865} (\bibinfo {year} {1996})}\BibitemShut {NoStop}%
\bibitem [{\citenamefont {Hamann}\ \emph {et~al.}(1979)\citenamefont {Hamann},
  \citenamefont {Schlüter},\ and\ \citenamefont
  {Chiang}}]{hamann_norm-conserving_1979}%
  \BibitemOpen
  \bibfield  {author} {\bibinfo {author} {\bibfnamefont {D.~R.}\ \bibnamefont
  {Hamann}}, \bibinfo {author} {\bibfnamefont {M.}~\bibnamefont {Schlüter}}, \
  and\ \bibinfo {author} {\bibfnamefont {C.}~\bibnamefont {Chiang}},\ }\href
  {\doibase 10.1103/PhysRevLett.43.1494} {\bibfield  {journal} {\bibinfo
  {journal} {Phys Rev Lett}\ }\textbf {\bibinfo {volume} {43}},\ \bibinfo
  {pages} {1494} (\bibinfo {year} {1979})}\BibitemShut {NoStop}%
\bibitem [{\citenamefont {Vanderbilt}(1985)}]{vanderbilt_optimally_1985}%
  \BibitemOpen
  \bibfield  {author} {\bibinfo {author} {\bibfnamefont {D.}~\bibnamefont
  {Vanderbilt}},\ }\href {\doibase 10.1103/PhysRevB.32.8412} {\bibfield
  {journal} {\bibinfo  {journal} {Phys Rev B}\ }\textbf {\bibinfo {volume}
  {32}},\ \bibinfo {pages} {8412} (\bibinfo {year} {1985})}\BibitemShut
  {NoStop}%
\bibitem [{\citenamefont {Bussi}\ \emph {et~al.}(2007)\citenamefont {Bussi},
  \citenamefont {Donadio},\ and\ \citenamefont
  {Parrinello}}]{bussi_canonical_2007}%
  \BibitemOpen
  \bibfield  {author} {\bibinfo {author} {\bibfnamefont {G.}~\bibnamefont
  {Bussi}}, \bibinfo {author} {\bibfnamefont {D.}~\bibnamefont {Donadio}}, \
  and\ \bibinfo {author} {\bibfnamefont {M.}~\bibnamefont {Parrinello}},\
  }\href {\doibase 10.1063/1.2408420} {\bibfield  {journal} {\bibinfo
  {journal} {J Chem Phys}\ }\textbf {\bibinfo {volume} {126}},\ \bibinfo
  {pages} {014101} (\bibinfo {year} {2007})}\BibitemShut {NoStop}%
\bibitem [{\citenamefont {Zhuang}\ \emph {et~al.}(2020)\citenamefont {Zhuang},
  \citenamefont {Ye}, \citenamefont {Pan},\ and\ \citenamefont
  {Li}}]{LinZhuang-43101}%
  \BibitemOpen
  \bibfield  {author} {\bibinfo {author} {\bibfnamefont {L.}~\bibnamefont
  {Zhuang}}, \bibinfo {author} {\bibfnamefont {Q.}~\bibnamefont {Ye}}, \bibinfo
  {author} {\bibfnamefont {D.}~\bibnamefont {Pan}}, \ and\ \bibinfo {author}
  {\bibfnamefont {X.-Z.}\ \bibnamefont {Li}},\ }\href {\doibase
  10.1088/0256-307X/37/4/043101} {\bibfield  {journal} {\bibinfo  {journal}
  {Chinese Phys Lett}\ }\textbf {\bibinfo {volume} {37}},\ \bibinfo {eid}
  {043101} (\bibinfo {year} {2020})}\BibitemShut {NoStop}%
\bibitem [{\citenamefont {Plimpton}(1995)}]{plimpton_fast_1995}%
  \BibitemOpen
  \bibfield  {author} {\bibinfo {author} {\bibfnamefont {S.}~\bibnamefont
  {Plimpton}},\ }\href {\doibase https://doi.org/10.1006/jcph.1995.1039}
  {\bibfield  {journal} {\bibinfo  {journal} {J Comput Phys}\ }\textbf
  {\bibinfo {volume} {117}},\ \bibinfo {pages} {1 } (\bibinfo {year}
  {1995})}\BibitemShut {NoStop}%
\bibitem [{\citenamefont {Wang}\ \emph {et~al.}(2018)\citenamefont {Wang},
  \citenamefont {Zhang}, \citenamefont {Han},\ and\ \citenamefont
  {E}}]{WANG2018178}%
  \BibitemOpen
  \bibfield  {author} {\bibinfo {author} {\bibfnamefont {H.}~\bibnamefont
  {Wang}}, \bibinfo {author} {\bibfnamefont {L.}~\bibnamefont {Zhang}},
  \bibinfo {author} {\bibfnamefont {J.}~\bibnamefont {Han}}, \ and\ \bibinfo
  {author} {\bibfnamefont {W.}~\bibnamefont {E}},\ }\href {\doibase
  https://doi.org/10.1016/j.cpc.2018.03.016} {\bibfield  {journal} {\bibinfo
  {journal} {Comput Phys Commun}\ }\textbf {\bibinfo {volume} {228}},\ \bibinfo
  {pages} {178 } (\bibinfo {year} {2018})}\BibitemShut {NoStop}%
\bibitem [{\citenamefont {Sun}\ \emph {et~al.}(2015)\citenamefont {Sun},
  \citenamefont {Ruzsinszky},\ and\ \citenamefont {Perdew}}]{sun2015strongly}%
  \BibitemOpen
  \bibfield  {author} {\bibinfo {author} {\bibfnamefont {J.}~\bibnamefont
  {Sun}}, \bibinfo {author} {\bibfnamefont {A.}~\bibnamefont {Ruzsinszky}}, \
  and\ \bibinfo {author} {\bibfnamefont {J.~P.}\ \bibnamefont {Perdew}},\
  }\href@noop {} {\bibfield  {journal} {\bibinfo  {journal} {Phys Rev Lett}\
  }\textbf {\bibinfo {volume} {115}},\ \bibinfo {pages} {036402} (\bibinfo
  {year} {2015})}\BibitemShut {NoStop}%
\bibitem [{\citenamefont {Abadi}\ \emph {et~al.}(2015)\citenamefont {Abadi},
  \citenamefont {Agarwal}, \citenamefont {Barham}, \citenamefont {Brevdo},
  \citenamefont {Chen}, \citenamefont {Citro}, \citenamefont {Corrado},
  \citenamefont {Davis}, \citenamefont {Dean}, \citenamefont {Devin},
  \citenamefont {Ghemawat}, \citenamefont {Goodfellow}, \citenamefont {Harp},
  \citenamefont {Irving}, \citenamefont {Isard}, \citenamefont {Jia},
  \citenamefont {Jozefowicz}, \citenamefont {Kaiser}, \citenamefont {Kudlur},
  \citenamefont {Levenberg}, \citenamefont {Man\'{e}}, \citenamefont {Monga},
  \citenamefont {Moore}, \citenamefont {Murray}, \citenamefont {Olah},
  \citenamefont {Schuster}, \citenamefont {Shlens}, \citenamefont {Steiner},
  \citenamefont {Sutskever}, \citenamefont {Talwar}, \citenamefont {Tucker},
  \citenamefont {Vanhoucke}, \citenamefont {Vasudevan}, \citenamefont
  {Vi\'{e}gas}, \citenamefont {Vinyals}, \citenamefont {Warden}, \citenamefont
  {Wattenberg}, \citenamefont {Wicke}, \citenamefont {Yu},\ and\ \citenamefont
  {Zheng}}]{tensorflow2015-whitepaper}%
  \BibitemOpen
  \bibfield  {author} {\bibinfo {author} {\bibfnamefont {M.}~\bibnamefont
  {Abadi}}, \bibinfo {author} {\bibfnamefont {A.}~\bibnamefont {Agarwal}},
  \bibinfo {author} {\bibfnamefont {P.}~\bibnamefont {Barham}}, \bibinfo
  {author} {\bibfnamefont {E.}~\bibnamefont {Brevdo}}, \bibinfo {author}
  {\bibfnamefont {Z.}~\bibnamefont {Chen}}, \bibinfo {author} {\bibfnamefont
  {C.}~\bibnamefont {Citro}}, \bibinfo {author} {\bibfnamefont {G.~S.}\
  \bibnamefont {Corrado}}, \bibinfo {author} {\bibfnamefont {A.}~\bibnamefont
  {Davis}}, \bibinfo {author} {\bibfnamefont {J.}~\bibnamefont {Dean}},
  \bibinfo {author} {\bibfnamefont {M.}~\bibnamefont {Devin}}, \bibinfo
  {author} {\bibfnamefont {S.}~\bibnamefont {Ghemawat}}, \bibinfo {author}
  {\bibfnamefont {I.}~\bibnamefont {Goodfellow}}, \bibinfo {author}
  {\bibfnamefont {A.}~\bibnamefont {Harp}}, \bibinfo {author} {\bibfnamefont
  {G.}~\bibnamefont {Irving}}, \bibinfo {author} {\bibfnamefont
  {M.}~\bibnamefont {Isard}}, \bibinfo {author} {\bibfnamefont
  {Y.}~\bibnamefont {Jia}}, \bibinfo {author} {\bibfnamefont {R.}~\bibnamefont
  {Jozefowicz}}, \bibinfo {author} {\bibfnamefont {L.}~\bibnamefont {Kaiser}},
  \bibinfo {author} {\bibfnamefont {M.}~\bibnamefont {Kudlur}}, \bibinfo
  {author} {\bibfnamefont {J.}~\bibnamefont {Levenberg}}, \bibinfo {author}
  {\bibfnamefont {D.}~\bibnamefont {Man\'{e}}}, \bibinfo {author}
  {\bibfnamefont {R.}~\bibnamefont {Monga}}, \bibinfo {author} {\bibfnamefont
  {S.}~\bibnamefont {Moore}}, \bibinfo {author} {\bibfnamefont
  {D.}~\bibnamefont {Murray}}, \bibinfo {author} {\bibfnamefont
  {C.}~\bibnamefont {Olah}}, \bibinfo {author} {\bibfnamefont {M.}~\bibnamefont
  {Schuster}}, \bibinfo {author} {\bibfnamefont {J.}~\bibnamefont {Shlens}},
  \bibinfo {author} {\bibfnamefont {B.}~\bibnamefont {Steiner}}, \bibinfo
  {author} {\bibfnamefont {I.}~\bibnamefont {Sutskever}}, \bibinfo {author}
  {\bibfnamefont {K.}~\bibnamefont {Talwar}}, \bibinfo {author} {\bibfnamefont
  {P.}~\bibnamefont {Tucker}}, \bibinfo {author} {\bibfnamefont
  {V.}~\bibnamefont {Vanhoucke}}, \bibinfo {author} {\bibfnamefont
  {V.}~\bibnamefont {Vasudevan}}, \bibinfo {author} {\bibfnamefont
  {F.}~\bibnamefont {Vi\'{e}gas}}, \bibinfo {author} {\bibfnamefont
  {O.}~\bibnamefont {Vinyals}}, \bibinfo {author} {\bibfnamefont
  {P.}~\bibnamefont {Warden}}, \bibinfo {author} {\bibfnamefont
  {M.}~\bibnamefont {Wattenberg}}, \bibinfo {author} {\bibfnamefont
  {M.}~\bibnamefont {Wicke}}, \bibinfo {author} {\bibfnamefont
  {Y.}~\bibnamefont {Yu}}, \ and\ \bibinfo {author} {\bibfnamefont
  {X.}~\bibnamefont {Zheng}},\ }\href {https://www.tensorflow.org/} {\enquote
  {\bibinfo {title} {{TensorFlow}: Large-scale machine learning on
  heterogeneous systems},}\ } (\bibinfo {year} {2015}),\ \bibinfo {note}
  {software available from tensorflow.org}\BibitemShut {NoStop}%
\bibitem [{\citenamefont {Nos{\'e}}(1984)}]{nose1984unified}%
  \BibitemOpen
  \bibfield  {author} {\bibinfo {author} {\bibfnamefont {S.}~\bibnamefont
  {Nos{\'e}}},\ }\href@noop {} {\bibfield  {journal} {\bibinfo  {journal} {J
  Phys Chem}\ }\textbf {\bibinfo {volume} {81}},\ \bibinfo {pages} {511}
  (\bibinfo {year} {1984})}\BibitemShut {NoStop}%
\bibitem [{\citenamefont {Hoover}(1985)}]{hoover1985canonical}%
  \BibitemOpen
  \bibfield  {author} {\bibinfo {author} {\bibfnamefont {W.~G.}\ \bibnamefont
  {Hoover}},\ }\href@noop {} {\bibfield  {journal} {\bibinfo  {journal} {Phys
  Rev A}\ }\textbf {\bibinfo {volume} {31}},\ \bibinfo {pages} {1695} (\bibinfo
  {year} {1985})}\BibitemShut {NoStop}%
\bibitem [{\citenamefont {Berendsen}\ \emph {et~al.}(1987)\citenamefont
  {Berendsen}, \citenamefont {Grigera},\ and\ \citenamefont
  {Straatsma}}]{berendsen_missing_1987}%
  \BibitemOpen
  \bibfield  {author} {\bibinfo {author} {\bibfnamefont {H.~J.~C.}\
  \bibnamefont {Berendsen}}, \bibinfo {author} {\bibfnamefont {J.~R.}\
  \bibnamefont {Grigera}}, \ and\ \bibinfo {author} {\bibfnamefont {T.~P.}\
  \bibnamefont {Straatsma}},\ }\href {\doibase 10.1021/j100308a038} {\bibfield
  {journal} {\bibinfo  {journal} {J Phys Chem}\ }\textbf {\bibinfo {volume}
  {91}},\ \bibinfo {pages} {6269} (\bibinfo {year} {1987})}\BibitemShut
  {NoStop}%
\bibitem [{\citenamefont {Abascal}\ and\ \citenamefont
  {Vega}(2005)}]{doi:10.1063/1.2121687}%
  \BibitemOpen
  \bibfield  {author} {\bibinfo {author} {\bibfnamefont {J.~L.~F.}\
  \bibnamefont {Abascal}}\ and\ \bibinfo {author} {\bibfnamefont
  {C.}~\bibnamefont {Vega}},\ }\href {\doibase 10.1063/1.2121687} {\bibfield
  {journal} {\bibinfo  {journal} {J Chem Phys}\ }\textbf {\bibinfo {volume}
  {123}},\ \bibinfo {pages} {234505} (\bibinfo {year} {2005})}\BibitemShut
  {NoStop}%
\bibitem [{\citenamefont {Hess}\ \emph {et~al.}(2008)\citenamefont {Hess},
  \citenamefont {Kutzner}, \citenamefont {van~der Spoel},\ and\ \citenamefont
  {Lindahl}}]{hess_gromacs_2008}%
  \BibitemOpen
  \bibfield  {author} {\bibinfo {author} {\bibfnamefont {B.}~\bibnamefont
  {Hess}}, \bibinfo {author} {\bibfnamefont {C.}~\bibnamefont {Kutzner}},
  \bibinfo {author} {\bibfnamefont {D.}~\bibnamefont {van~der Spoel}}, \ and\
  \bibinfo {author} {\bibfnamefont {E.}~\bibnamefont {Lindahl}},\ }\href
  {\doibase 10.1021/ct700301q} {\bibfield  {journal} {\bibinfo  {journal} {J
  Chem Theory Comput}\ }\textbf {\bibinfo {volume} {4}},\ \bibinfo {pages}
  {435} (\bibinfo {year} {2008})}\BibitemShut {NoStop}%
\bibitem [{\citenamefont {Gereben}\ and\ \citenamefont
  {Pusztai}(2011)}]{gereben_accurate_2011}%
  \BibitemOpen
  \bibfield  {author} {\bibinfo {author} {\bibfnamefont {O.}~\bibnamefont
  {Gereben}}\ and\ \bibinfo {author} {\bibfnamefont {L.}~\bibnamefont
  {Pusztai}},\ }\href {\doibase 10.1016/j.cplett.2011.02.064} {\bibfield
  {journal} {\bibinfo  {journal} {Chem Phys Lett}\ }\textbf {\bibinfo {volume}
  {507}},\ \bibinfo {pages} {80} (\bibinfo {year} {2011})}\BibitemShut
  {NoStop}%
\bibitem [{\citenamefont {Darden}\ \emph {et~al.}(1993)\citenamefont {Darden},
  \citenamefont {York},\ and\ \citenamefont {Pedersen}}]{darden1993particle}%
  \BibitemOpen
  \bibfield  {author} {\bibinfo {author} {\bibfnamefont {T.}~\bibnamefont
  {Darden}}, \bibinfo {author} {\bibfnamefont {D.}~\bibnamefont {York}}, \ and\
  \bibinfo {author} {\bibfnamefont {L.}~\bibnamefont {Pedersen}},\ }\href@noop
  {} {\bibfield  {journal} {\bibinfo  {journal} {J Chem Phys}\ }\textbf
  {\bibinfo {volume} {98}},\ \bibinfo {pages} {10089} (\bibinfo {year}
  {1993})}\BibitemShut {NoStop}%
\bibitem [{\citenamefont {Zhang}\ and\ \citenamefont
  {Duan}(2005)}]{zhang_prediction_2005}%
  \BibitemOpen
  \bibfield  {author} {\bibinfo {author} {\bibfnamefont {Z.}~\bibnamefont
  {Zhang}}\ and\ \bibinfo {author} {\bibfnamefont {Z.}~\bibnamefont {Duan}},\
  }\href {\doibase https://doi.org/10.1016/j.pepi.2004.11.003} {\bibfield
  {journal} {\bibinfo  {journal} {Phys Earth Planet Inter}\ }\textbf {\bibinfo
  {volume} {149}},\ \bibinfo {pages} {335 } (\bibinfo {year}
  {2005})}\BibitemShut {NoStop}%
\end{thebibliography}%
 
\newpage

\begin{figure}
\centering
\vspace{5mm}
\includegraphics[width=0.6 \textwidth]{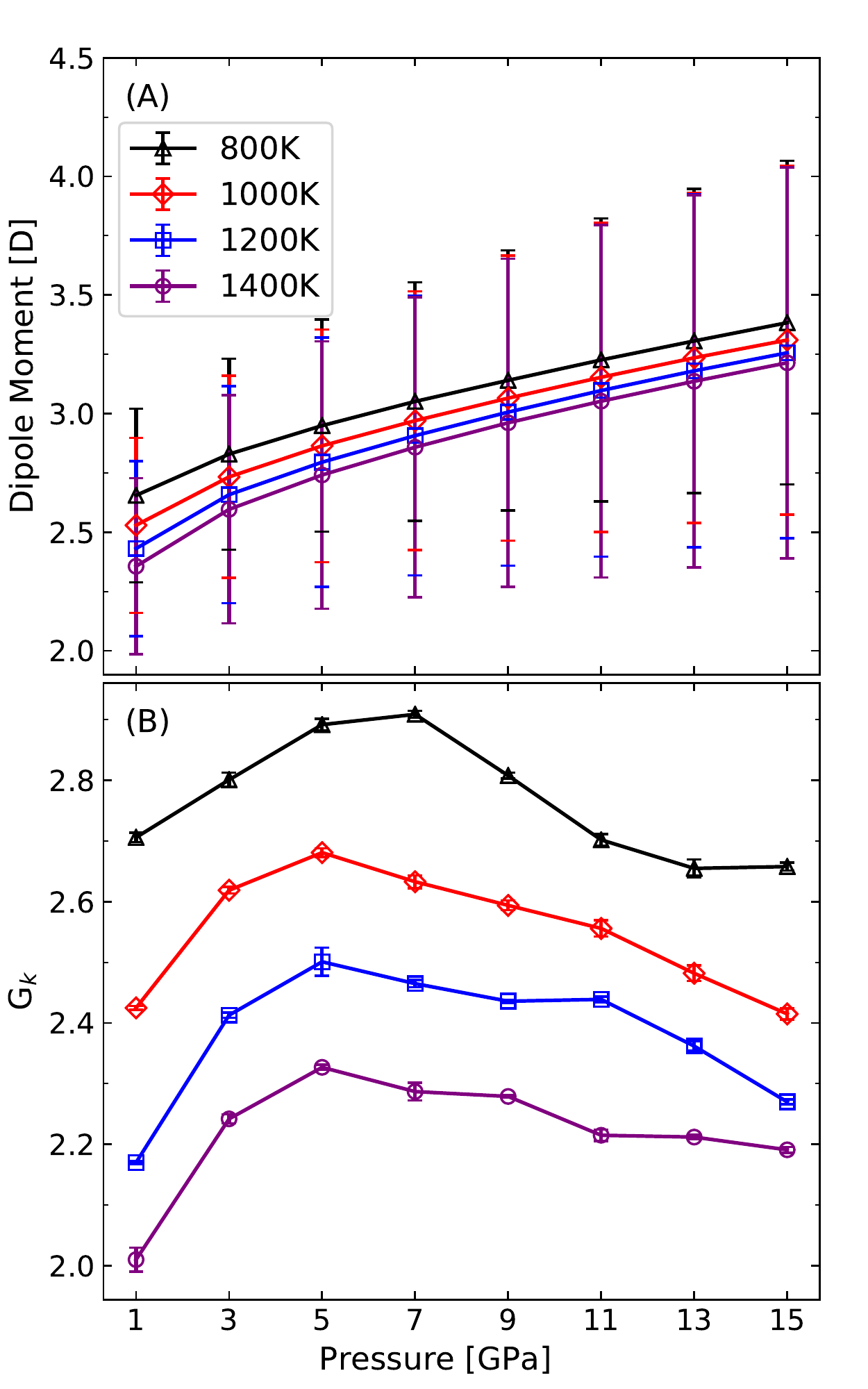}
\caption{Molecular dipole moments and the Kirkwood g-factor, $G_k$, of water at high P-T conditions. (A) The average molecular dipole moment and standard derivation as a function of pressure. (B) $G_k$ as a function of pressure. }
\label{G-factor}
\end{figure}

\begin{figure}
\centering
\vspace{5mm}
\includegraphics[width=0.6 \textwidth]{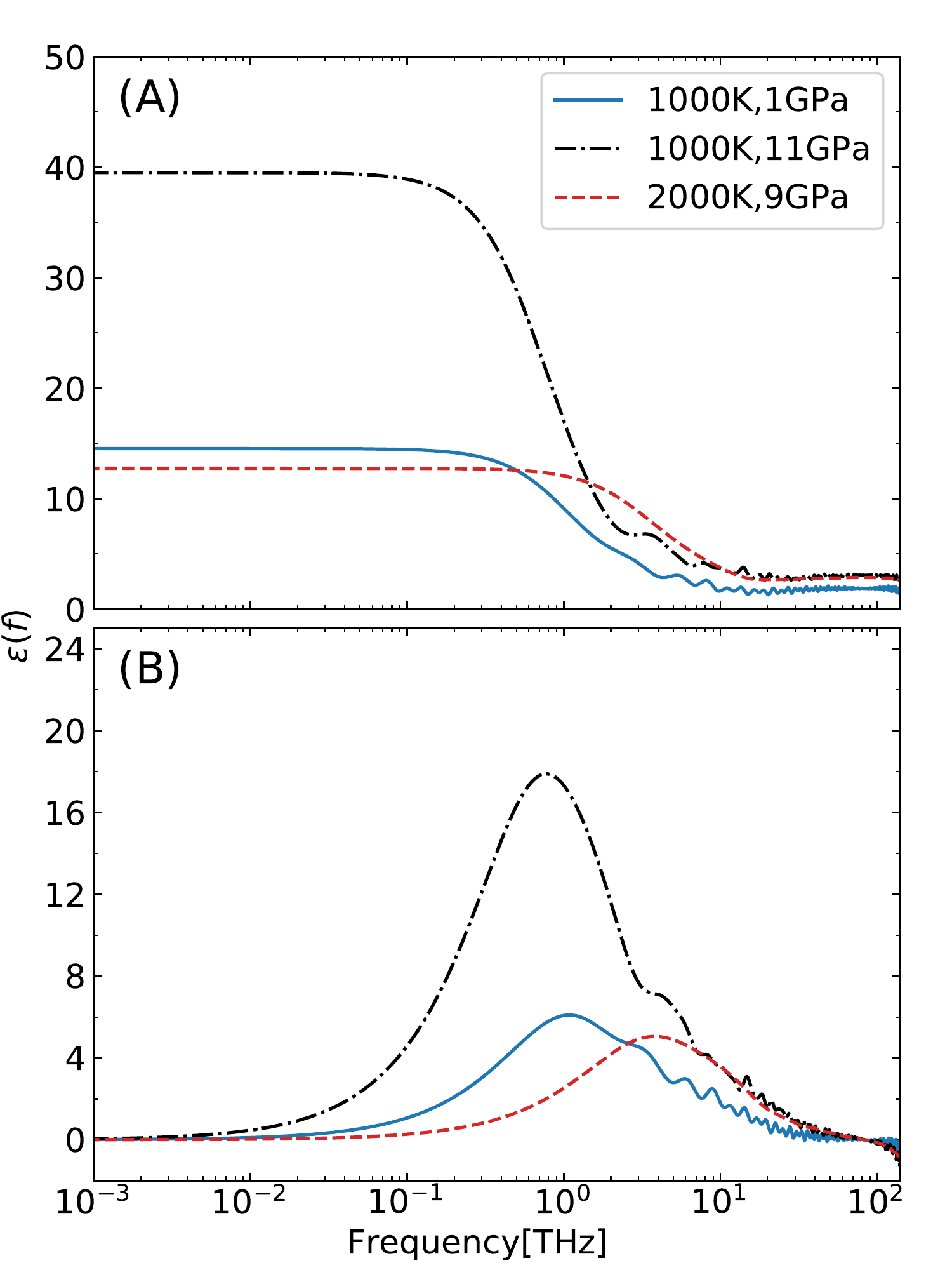}
\caption{Frequency-dependent dielectric constant of water, $\epsilon (f)$, at high P-T conditions. The real and imaginary parts of $\epsilon (f)$ in the microwave range are shown in (A) and (B), respectively ($\omega = 2\pi f $). }
\label{fig:epsw}
\end{figure}

\end{document}

\title{Supplementary Materials}
\author{Rui Hou}
\affiliation{Department of Physics, Hong Kong University of Science and Technology, Hong Kong, China}
\author{Yuhui Quan}
\affiliation{Department of Physics, Hong Kong University of Science and Technology, Hong Kong, China}
\author{Ding Pan}
\email{dingpan@ust.hk}
\affiliation{Department of Physics, Hong Kong University of Science and Technology, Hong Kong, China}
\affiliation{Department of Chemistry, Hong Kong University of Science and Technology, Hong Kong, China}
\affiliation{HKUST Fok Ying Tung Research Institute, Guangzhou, China}

\date{\today}

\maketitle

\subsection{First principles molecular dynamics simulations (FPMD)}

We performed FPMD simulations in the Born–Oppenheimer approximation using the Qbox package (v.1.54.2 \cite{gygi_architecture_2008} http://qboxcode.org/). The MD time step was 0.24 fs. We used the PBE exchange-correlation functional \cite{Perdew1996} and norm-conserving pseudopotentials \cite{hamann_norm-conserving_1979, vanderbilt_optimally_1985} (pseudopotential table, http://fpmd.ucdavis.edu/potentials/). The kinetic energy cutoff was 85 Ry, and was increased to 220 Ry to calculate pressure. The cubic simulation box with periodic boundary conditions contains 128 heavy water molecules (D$_2$O).  The density of water mentioned in the main text was calculated for H$_2$O. We used the stochastic velocity rescaling thermostat ($\tau$ = 24.2 fs) to control temperature \cite{bussi_canonical_2007}. We started FPMD simulations after SPC/E force field simulations, and carried out for at least 20 ps for each P-T condition.

\subsection{Neural network molecular dynamics (NNMD)}
We performed MD simulations with the neural network (NN) force field \cite{LinZhuang-43101} using the Lammps package \cite{plimpton_fast_1995} with the DeepMD-kit module \cite{WANG2018178}. The NN force field was trained using the data from FPMD simulations with the SCAN functional \cite{sun2015strongly}. The machine learning package is TensorFlow (version 1.14.0) \cite{tensorflow2015-whitepaper}. The MD time step was 0.12 fs. 
We used the Nosé–Hoover thermostat to control temperature, and the T damping parameter was 50 fs \cite{nose1984unified, hoover1985canonical}. 
In Fig. 2 and Table SI, the simulation box contains 128 H$_2$O molecules. Each MD trajectory is about 112 ps and the equilibration time is 12 ps. 
In Fig. 3 and Table SII, the simulation box contains 256 H$_2$O molecules.
Each MD trajectory is at least 2.1 ns.
We calculated molecular dipole moments every 200 MD steps. 

\subsection{Molecular dynamics simulations using the SPC/E and TIP4P/2005 models}
We performed SPC/E \cite{berendsen_missing_1987} and TIP4P/2005 \cite{doi:10.1063/1.2121687}  molecular dynamics simulations using the Gromacs package (version: 2019.4) \cite{hess_gromacs_2008}. The MD time step was 1 fs. We used the stochastic velocity rescaling thermostate ($\tau$ = 100 fs) to control temperature \cite{bussi_canonical_2007}. 

Each MD trajectory exceeds 8 ns \cite{gereben_accurate_2011} and the equilibration time is more than 1 ns. The real-part electrostatics and Lennard–Jones (LJ) potentials have a radius of 1.56 nm ( P $\sim$ 1GPa, T = 1000K), 1.32 nm (P $\sim$ 10GPa, T = 1000K), 1.47 nm (P $\sim$ 5GPa, T = 2000K), or 1.38 nm (P $\sim$ 10GPa, T = 2000K). The long-range electrostatics interactions were calculated using the particle mesh Ewald (PME) method \cite{darden1993particle}. The simulation box contains 1024 water molecules. 

\newpage

\begin{figure}
\centering
\vspace{5mm}
\includegraphics[width=0.8 \textwidth]{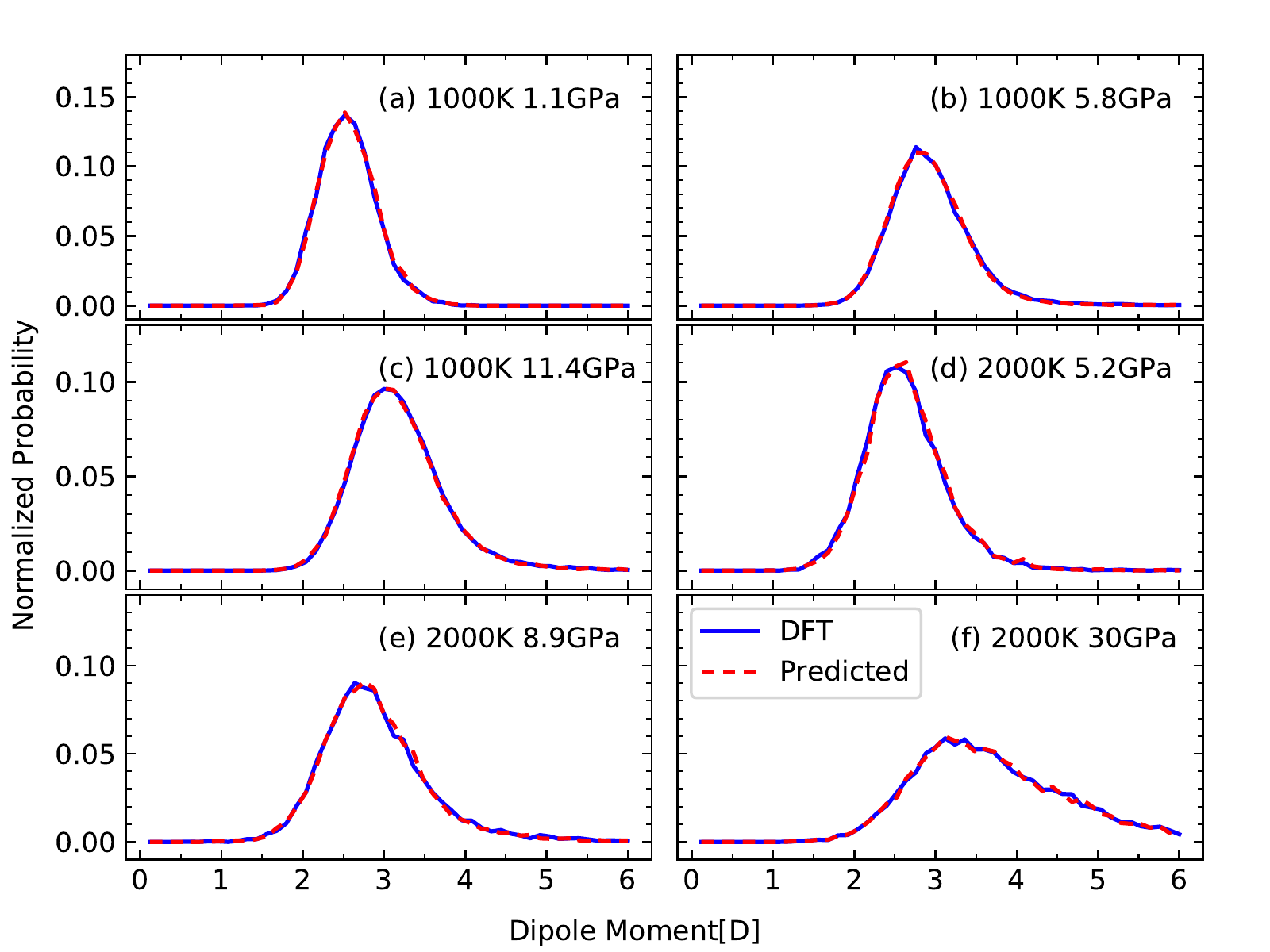}
\caption{Distribution of dipole magnitude of water molecules under high P-T conditions. The neural network dipole model (red dashed lines) and DFT (blue solid lines) results are compared. }
\label{fig:dipole}
\end{figure}

\begin{figure}
  \centering
  \vspace{5mm}
  \includegraphics[width=0.8 \textwidth]{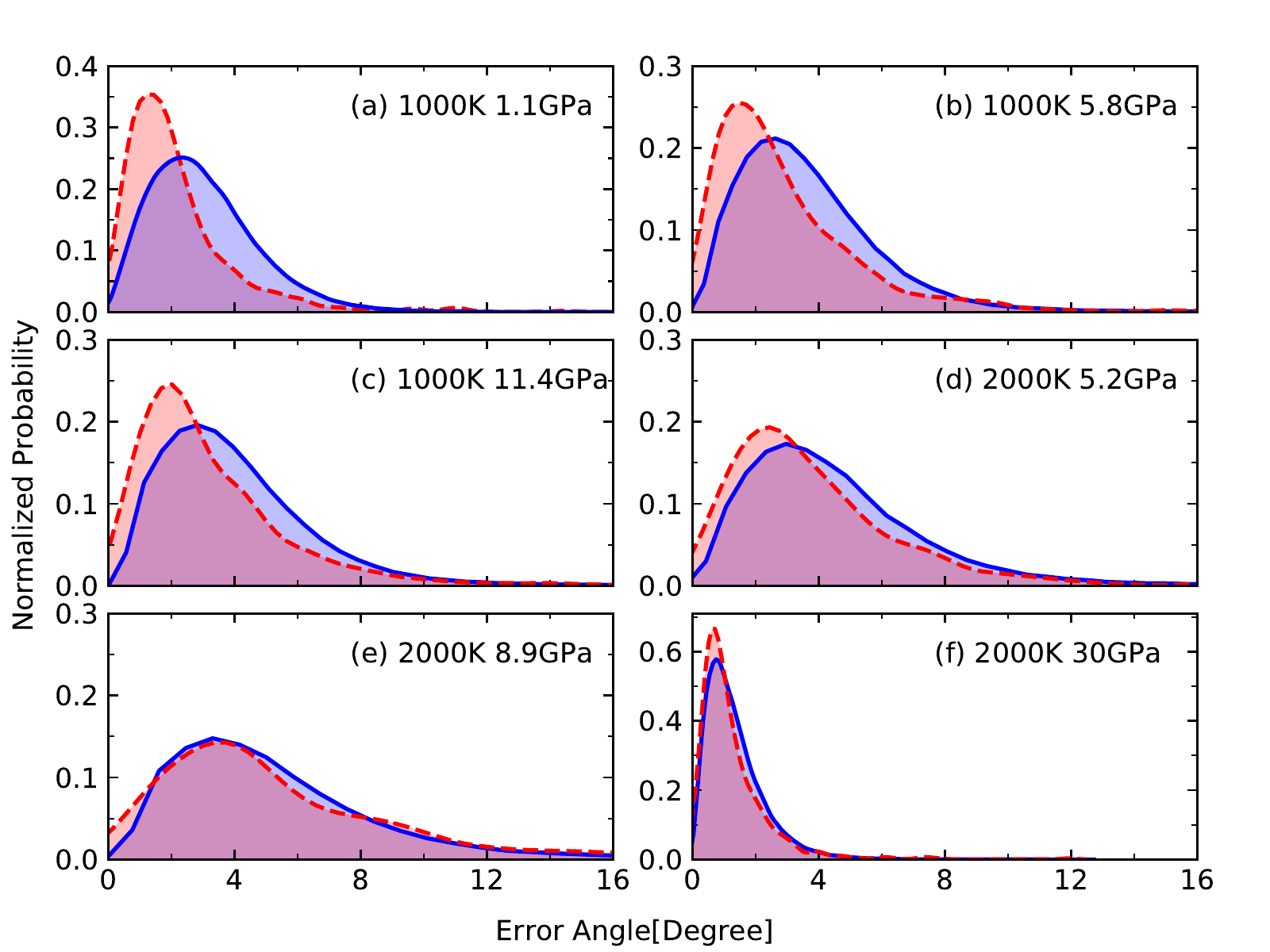}
  \caption{Distribution of the angle between dipole moment vectors calculated by DFT and the neural network dipole model. The blue solid and red dashed curves show the results for the individual dipole moments of water molecules and the total dipole moment of simulation boxes, respectively.}
  \label{fig:error}
  \end{figure}

  \begin{table}[H]
    \centering
    \caption{Static dielectric constant of water obtained from several force fields. The asterisk (*) denotes that the trajectory is from the corresponding force field, while molecular dipoles are calculated using the neural network dipole model. Pressures are calculated by DFT. Standard deviations are reported within parentheses.
    }
    \label{table:pt}
    \begin{tabular}{|c|c|c|c|c|c|}
      \hline 
      T (K)  & \multicolumn{3}{|c|}{1000} & \multicolumn{2}{|c|}{2000} \\ \hline 
      P (GPa)  & 1.1(0.2) & 5.8(0.6) & 11.4(0.4) & 5.2(0.5) & 8.9(1.0) \\ \hline 
      $\rho (g/cm^3)$       & 0.88 & 1.32 & 1.57 & 1.13 & 1.36  \\ \hline 
      $\epsilon_{\infty}$  &  1.761 & 2.175 & 2.411 & 2.049 & 2.294  \\ \hline 
      DFT           & 15.11(1.32) & 29.91(3.39) & 39.61(3.25) & 9.89(0.55) & 13.57(0.18) \\ \hline 
      DFT*       & 15.02(1.30) & 29.82(3.36) & 39.72(3.25) & 9.87(0.55) & 13.51(0.18) \\ \hline 
      TIP4P/2005     & 11.28(0.32) & 18.36(0.15) & 22.50(0.14) & 6.47(0.07) & 8.73(0.11) \\ \hline
      TIP4P/2005*   & 14.48(0.36) & 27.35(0.35) & 36.77(0.45) & 9.39(0.03) & 12.36(0.13) \\ \hline
      SPC/E          & 12.39(0.13) & 21.83(0.16) & 27.70(0.18) & 6.96(0.08) & 10.04(0.15) \\ \hline
      SPC/E*      & 15.66(0.14) & 32.48(0.41) & 45.63(0.40) & 9.71(0.08) & 13.58(0.15) \\ \hline
      NNMD*    & 14.67(0.37) & 31.06(0.67) & 40.19(1.40) & 10.04(0.19)& 13.33(0.13) \\ \hline
      \end{tabular}
    \end{table}

        \begin{table}[H]
      \caption{Static and electronic dielectric constants of water predicted by the neural network dipole model and NNMD simulations. The equation of state of water is from Ref. \cite{zhang_prediction_2005}. Standard deviations are reported within parentheses.}
      \begin{tabular}{|c|c|c|c|c|c|c|c|c|c|} \hline
      T/K  & P/GPa & $\rho$/g$\cdot$ cm$^{-3}$ & $\epsilon_{\infty}$ & $\epsilon_0$ & T/K  & P/GPa & $\rho$/g$\cdot$ cm$^{-3}$ & $\epsilon_{\infty}$ & $\epsilon_0$ \\ \hline
      % T/K  & P/GPa & cc\cdot mol^{-1} &  & T/K  & P/GPa & /cc\cdot mol^{-1} & \\ \hline
      800  & 1     & 0.99   & 1.84 & 25.89(0.07)     & 1000 & 1     & 0.90   & 1.76 & 15.99(0.02)     \\ \hline
      800  & 3     & 1.26   & 2.09 & 37.93(0.15)     & 1000 & 3     & 1.20   & 2.04 & 25.84(0.05)    \\ \hline
      800  & 5     & 1.41   & 2.22 & 47.08(0.15)     & 1000 & 5     & 1.35   & 2.18 & 32.38(0.08)     \\ \hline
      800  & 7     & 1.51   & 2.32 & 54.27(0.10)     & 1000 & 7     & 1.46   & 2.29 & 36.77(0.14)    \\ \hline
      800  & 9     & 1.60   & 2.39 & 58.53(0.10)     & 1000 & 9     & 1.55   & 2.37 & 40.70(0.12)     \\ \hline
      800  & 11    & 1.67   & 2.45 & 62.21(0.21)     & 1000 & 11    & 1.62   & 2.43 & 44.27(0.22)     \\ \hline
      800  & 13    & 1.73   & 2.51 & 66.31(0.36)     & 1000 & 13    & 1.69   & 2.49 & 46.90(0.23)    \\ \hline
      800  & 15    & 1.79   & 2.56 & 71.59(0.17)     & 1000 & 15    & 1.74   & 2.54 & 49.28(0.18)     \\ \hline
      1200 & 1     & 0.82   & 1.69 & 10.63(0.01)     & 1400 & 1     & 0.76   & 1.62 & 7.73(0.06)      \\ \hline
      1200 & 3     & 1.14   & 2.00 & 18.47(0.03)     & 1400 & 3     & 1.09   & 1.95 & 13.91(0.04)     \\ \hline
      1200 & 5     & 1.30   & 2.14 & 23.69(0.20)     & 1400 & 5     & 1.26   & 2.12 & 18.07(0.03)     \\ \hline
      1200 & 7     & 1.42   & 2.26 & 27.22(0.06)     & 1400 & 7     & 1.38   & 2.24 & 20.85(0.12)     \\ \hline
      1200 & 9     & 1.51   & 2.34 & 30.37(0.03)     & 1400 & 9     & 1.47   & 2.31 & 23.53(0.11)    \\ \hline
      1200 & 11    & 1.58   & 2.41 & 33.66(0.06)     & 1400 & 11    & 1.54   & 2.39 & 25.43(0.10)     \\ \hline
      1200 & 13    & 1.64   & 2.46 & 35.65(0.11)     & 1400 & 13    & 1.61   & 2.46 & 27.75(0.04)     \\ \hline
      1200 & 15    & 1.70   & 2.52 & 36.08(0.06)     & 1400 & 15    & 1.66   & 2.50 & 27.80(0.06)     \\ \hline       
      \end{tabular}
      \end{table}

\bibliography{SI-ref}